\documentclass[aps,prb,twocolumn,amsmath,amssymb,showpacs,superscriptaddress,notitlepage]{revtex4-1}
\usepackage{amsmath,amssymb,amsfonts,bm}
\usepackage{graphicx}
\usepackage{epstopdf}
\usepackage{dcolumn}
\usepackage{mathrsfs}
\usepackage{bbold}
\usepackage{dsfont}
\usepackage{dcolumn}
\usepackage[colorlinks=true,linkcolor=blue,citecolor=blue, urlcolor=blue,bookmarks=false]{hyperref}
\usepackage{changes}
\colorlet{Changes@Color}{red}

\begin{document}

\title{Quasicrystalline second-order topological semimetals}

\author{Rui Chen}
\affiliation{Department of Physics, Hubei University, Wuhan 430062, China}
\author{Bin Zhou}
\affiliation{Department of Physics, Hubei University, Wuhan 430062, China}
\author{Dong-Hui Xu}\email[]{donghuixu@cqu.edu.cn}
\affiliation{Department of Physics, Chongqing University, Chongqing 400044, China}
\affiliation{Chongqing Key Laboratory for Strongly Coupled Physics, Chongqing University, Chongqing 400044, China}

\begin{abstract}
Three-dimensional higher-order topological semimetals in crystalline systems exhibit higher-order Fermi arcs on one-dimensional hinges, challenging the conventional bulk-boundary correspondence. However, the existence of higher-order Fermi arc states in aperiodic quasicrystalline systems remains uncertain. In this work, we present the emergence of three-dimensional quasicrystalline second-order topological semimetal phases by vertically stacking two-dimensional quasicrystalline second-order topological insulators. These quasicrystalline topological semimetal phases are protected by rotational symmetries forbidden in crystals, and are characterized by topological hinge Fermi arcs connecting fourfold degenerate Dirac-like points in the spectrum. Our findings reveal an intriguing class of higher-order topological phases in quasicrystalline systems, shedding light on their unique properties.
\end{abstract}
\maketitle
\section{Introduction}

\begin{figure}[h]
\centering
\includegraphics[width =1\columnwidth]{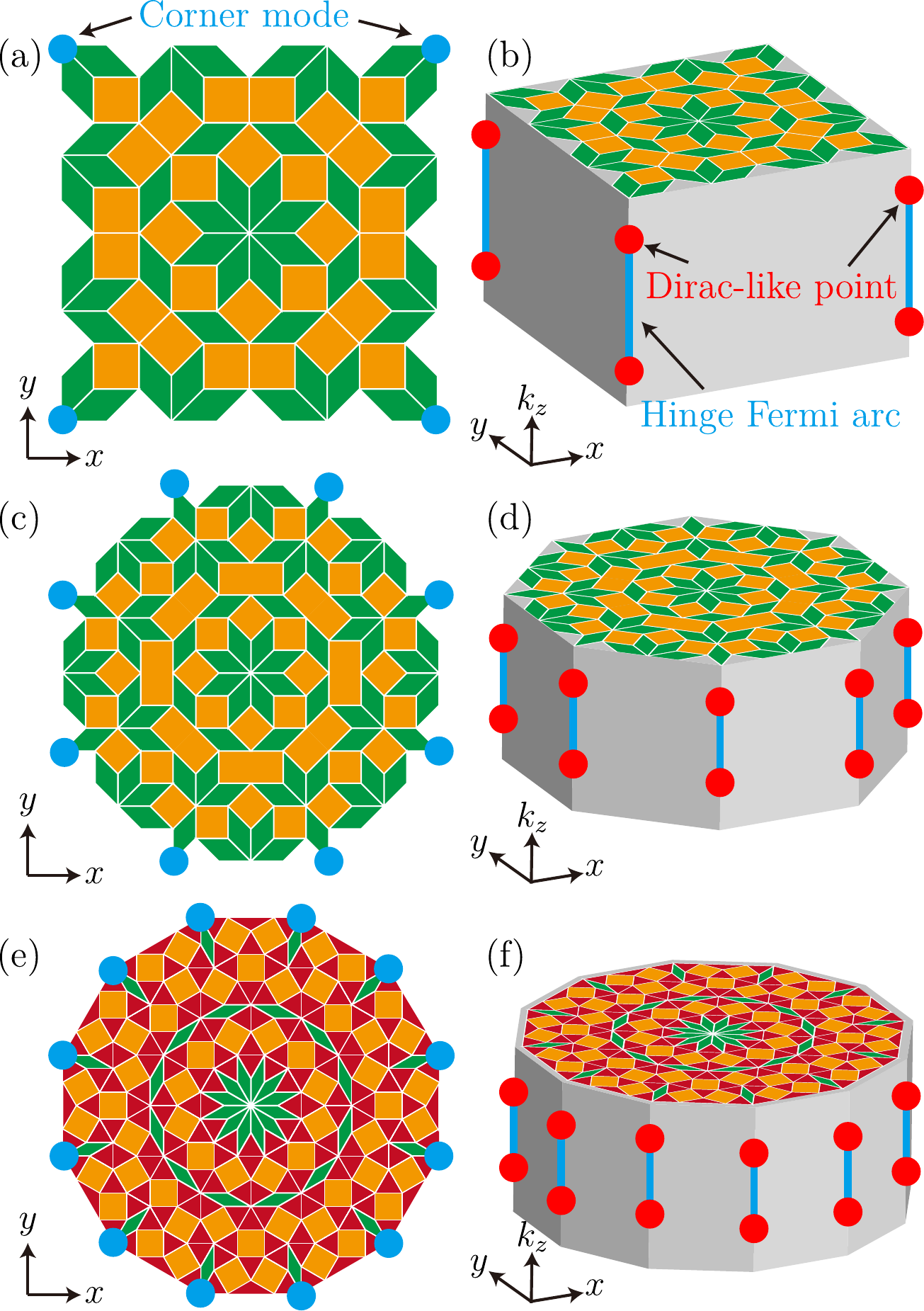}
\caption{Schematic illustrations of (a) AB-tiling square, (c) AB-tiling octagonal, and (e) Stampfli-tiling dodecagonal quasicrystals.
AB-tiling quasicrystal consists of two types of primitive tiles: square tiles (yellow) and rhombus tiles (green) with a small angle 45$^{\circ}$. Stampfli-tiling quasicrystal consists of three types of primitive tiles:
square tiles (yellow), regular triangle tiles (red), and rhombus tiles
(green) with a small angle 30$^{\circ}$. In (a, c, e), the side lengths (the white lines) of the polygons are taken as the length unit with $a=1$. 2D quasicrystalline SOTIs support (a) four, (c) eight, and (e) twelve rotation-symmetry-protected zero-energy corner modes (the cyan points). A simple stack of the 2D quasicrystalline SOTIs gives rise to 3D quasicrystalline SOTSMs with (b) four, (d) eight, and (f) twelve rotation-symmetry-protected hinge Fermi arc states (the cyan lines) connecting the Dirac-like points (the red points) in the spectrum.}
\label{fig_illustration}
\end{figure}

Symmetry-protected topological phases of matter have emerged as a major new theme in modern condensed-matter physics in the past nearly two decades. While the discovery of topological insulators initially sparked interest in this field, recent focus has shifted towards exploring higher-order topological insulators~\cite{Benalcazar2017Science,Schindler2018SciAdv,Langbehn17PRL,
Benalcazar17PRB,Song17PRL}. Unlike traditional topological insulators, higher-order topological insulators exhibit unconventional bulk-boundary correspondence, allowing for the existence of gapless boundary excitations of higher co-dimensions. For example, a second-order topological insulator (SOTI) in two dimensions hosts robust gapless boundary modes localized at its zero-dimensional corners, dubbed corner modes~\cite{Benalcazar2017Science}, while three-dimensional~(3D) SOTIs support gapless boundary modes confined to their one-dimensional hinges~\cite{Schindler2018SciAdv}. In addition to higher-order topological insulators, higher-order topological semimetals have also been identified. These semimetals, including higher-order Dirac semimetals and higher-order Weyl semimetals, exhibit exotic hinge Fermi arcs that connect the projected nodes on the hinges, distinguishing them from conventional Dirac and Weyl semimetals~\cite{LiCZ20PRL,Tyner21arXiv,
Nie22SB,Yuan21PRB,Wieder2020NC,Zeng2022FronPhys,Roy20PRR,LinMao18PRB,
ChenR21PRL,WangZM22arXiv,Wieder2020NC,Wang20PRL}.

Initially, topological phases were observed in crystalline materials. However, more recently, researchers have extended these phases to aperiodic quasicrystalline systems, which lack discrete translational symmetry~\cite{Tran2015PRB,Duncan20PRB,Li2020CPB,Jeon22PRB,He19PRB,Traverso22PRB,
	Huang18PRL,Spurrier20PRR,Bandres2016PRX,Poyhonen2018NatCom,Fulga2016PRL,Manna2022arXiv,
	Longhi19PRL,Zeng20PRB,ChenR20PRL,Varjas19PRL,Hua20PRB,PhysRevResearch.2.033071,Huang2021NanoLetter,Wang22PRL,Shi2022arXiv,Xiong2022arXiv,
	Ghadimi21PRB,Lv2021ComPhys,ChenR19PRB,Hua21PRB,Peng21PRB,Peng21PRB1,
	Verbin2013PRL,Apigo2019PRL,Shechtman1984PRL,Levine1984PRL,Biggs1990PRL,Pierce1993Science,Basov1994PRL,Pierce1994PRL,Cain2020PNAS}. The absence of translational symmetry allows for the presence of rotational symmetries that are prohibited in crystals. This property enables the existence of new topological phases without crystalline counterparts, such as two-dimensional (2D) SOTIs protected by eightfold~\cite{ChenR20PRL,Varjas19PRL} and twelvefold~\cite{Hua20PRB,PhysRevResearch.2.033071} rotational symmetries. Moreover, a 3D time-reversal symmetry (TRS) breaking gapless topological phase hosting Weyl-like points has been proposed in a quasicrystal stack of Chern insulators~\cite{Fonseca2022arXiv}.
	However, gapless phases with higher-order topology in quasicrystalline systems have yet to be discovered. This knowledge gap motivates us to explore the possibility of gapless quasicrystalline higher-order topological phases using a stacking approach with 2D quasicrystalline SOTIs. It has been demonstrated that stacking 2D topological materials provides a natural way of realizing 3D topological phases. This approach has been successful in achieving various topological phases, including Weyl semimetals~\cite{Burkov2011PRL}, axion insulators~\cite{Mogi17nm,Mogi17sa,Deng20sci,Gao2021Nature,ChenR21PRB}, hinged quantum spin Hall insulators~\cite{Ding20prbrc,Shumiya2022NatMat}, and high-Chern number quantum anomalous Hall insulators~\cite{Zhao2020Nature}.

In this work, we present the discovery of a quasicrystalline second-order topological semimetal (SOTSM) phase obtained by stacking 2D quasicrystalline SOTIs along the vertical direction~(Fig.~\ref{fig_illustration}). The distinctive feature of the quasicrystalline SOTSM is the presence of rotation-symmetry-protected topological hinge Fermi arcs that terminate at fourfold degenerate Dirac-like points in the spectrum. The $C_n^z$-symmetric quasicrystalline SOTSM can support $n$ topological hinge Fermi arcs (see the second column in Fig.~\ref{fig_illustration}), inheriting their topological nature from $C_n^z$-symmetric quasicrystalline SOTI hosting $n$ corner modes (see the first column in Fig.~\ref{fig_illustration}). The number $n$ can be four~[Figs.~\ref{fig_illustration}(a) and \ref{fig_illustration}(b)], as allowed in crystalline systems~\cite{Roy20PRR,LinMao18PRB,Yuan21PRB,Zeng2022FronPhys,Wieder2020NC,Wang20PRL}, but it can also be eight~[Figs.~\ref{fig_illustration}(c) and \ref{fig_illustration}(d)] and twelve~[Figs.~\ref{fig_illustration}(e) and \ref{fig_illustration}(f)], which are typically forbidden in crystalline systems. Furthermore, we present the phase diagram of the stacked systems and identify a 3D quasicrystalline SOTI phase in addition to the quasicrystalline SOTSM phase. Finally, we show that the disclination-induced bound states can further reveal the topological nature of the quasicrystalline SOTSM phase.

This work is organized as follows. We first give a simple review of 2D quasicrystalline SOTI in Sec.~\ref{review} and show a stack of it gives rise to the 3D quasicrystalline SOTSM phase with Dirac-like points in the spectrum in Sec.~\ref{SOTSM}. A detailed discussion on Dirac-like points is presented in Sec.~\ref{Sec_Gaplesspoint}. Subsequently, we illustrate the phase diagram of the stacked quasicrystalline system in Sec.~\ref{Phase} and investigate the disclination-induced bound state in Sec.~\ref{Defect}. We summarize our conclusions and discuss possible experimental
schemes for the quasicrystalline SOTSM phase in Sec.~\ref{discussion}. 

\section{Review of 2D quasicrystalline SOTIs}
\label{review}
2D quasicrystalline SOTIs had been proposed in
eightfold symmetric Ammann-Beenker-tiling (AB-tiling) quasicrystal~\cite{ChenR20PRL,Varjas19PRL} [Figs.~\ref{fig_illustration}(a) and~\ref{fig_illustration}(c)] and twelvefold symmetric Stampfli-tiling quasicrystal~\cite{Hua20PRB} [Fig.~\ref{fig_illustration}(e)]. The AB-tiling quasicrystal consists of two types of primitive tiles: square tiles (yellow) and rhombus tiles (green) with a small angle 45$^{\circ}$. The Stampfli-tiling quasicrystal consists of three types of primitive tiles:
square tiles (yellow), regular triangle tiles (red), and rhombus tiles (green) with a small angle 30$^{\circ}$.

In the tight-binding model, the lattice sites are placed on the vertices of each tile. The Hamiltonian of the 2D quasicrystalline SOTI contains two parts, $H(M)=H_\text{1st}(M)+H_\text{m}$~\cite{ChenR20PRL}. The first part denotes a 2D first-order topological insulator protected by TRS
\begin{eqnarray}
H_\text{1st}(M)&=&-\sum_{j\neq k}\frac{Z(r_{jk})}{2}\big[it_{1}\left( s _{3}\tau _{1}\cos\phi_{jk}+s _{0}\tau _{2}\sin\phi_{jk}\right)\nonumber\\
&+&t_{2}s _{0}\tau_{3}\big]  c_{j}^{\dag }c_{k}+\sum_{j} \left(M+2t_{2}\right)s _{0}\tau _{3} c_{j}^{\dag }c_{j},
\label{model1}
\end{eqnarray}
where $c^\dagger_{j\alpha}=(c^\dagger_{j\alpha\uparrow},c^\dagger_{j\alpha\downarrow})$ are electron creation operators at site $j$ with the orbital $\alpha$. $t_1$ and $t_2$ are hopping amplitudes, and $M$ denotes the Dirac mass, together with $t_2$, determining the first-order topology. $s_{1,2,3}$ and $\tau_{1,2,3}$ are the Pauli matrices acting on the spin and orbital spaces, respectively. $s_0$ is the $2\times 2$ identity matrix. $\phi_{jk}$ is the azimuthal angle of the bond between site $j$ and $k$ with respect to the horizontal direction. $Z\left( r_{jk}\right) = e^{1-r_{jk}/\xi}$ is the spatial decay factor of hopping amplitudes with the decay length $\xi$.
The second part is a TRS breaking Wilson mass term, which is
\begin{equation}
H_\text{m}(\eta)=g\sum_{j\neq k}\frac{Z(r_{jk})}{2}\cos\left( \eta\phi_{jk} \right) s _{1}\tau _{1}  c_{j}^{\dag }c_{k},
\end{equation}
where $g$ and $\eta$ describe the magnitude and varying period of the Wilson mass, respectively. $H_\text{m}(\eta)$ are responsible for higher-order topology~\cite{ChenR20PRL,Agarwala2020PRR}. In the subsequent calculations, we fix the side length of the tiles as $a=1$ (white lines connecting the vertices in Fig.~\ref{fig_illustration}) and $\xi=t_1=1$.

For $\eta=2,4,6$, the Wilson mass gives rise to the SOTI phases in quasicrystals hosting four, eight, and twelve corner modes protected by the combined symmetry $C_4^z U$~\cite{ChenR20PRL,Agarwala2020PRR}, $C_8^z U$~\cite{ChenR20PRL}, and $C_{12}^z U$~\cite{Hua20PRB}, respectively, where $C_n^z$ is the $n$-fold rotational operation, and $U$ could be the TRS operation $T=i s_2\tau_0 K$ or the mirror symmetry operation $m_z=s_3 \tau_0$. $K$ is the complex conjugation operator. The symmetry-protected eightfold and twelvefold corner modes, which are impossible in crystals~\cite{WangY2019PRB}, are distinguishing characteristics of the 2D quasicrystalline SOTIs. Additionally, these corner modes are pinned to zero energy due to the existence of particle-hole symmetry.

The emergence of the zero-energy corner modes can be simply understood as follows~\cite{Tao2022arXiv,Wang22PRL,ChenR20PRL}: $g$ opens a gap in the first-order topological edge states and then induces Wilson mass kinks near the boundary. If one corner mode $\left|\psi_c\right>$ appears at $\mathbf{r}_c$, where the Wilson mass flips the sign, then the $C_n^z U$ symmetry ensures that the number of corner modes is $n$. Because $C_n^z U\left|\psi_c\right>$ is also the eigenstate of the system, which is localized at another corner by rotating $\mathbf{r}_c$ by an angle of $2\pi/n$.

\section{3D quasicrystalline SOTSMs}
\label{SOTSM}
3D crystalline SOTSMs have been constructed by stacking 2D crystalline SOTIs along the vertical direction~\cite{LiCZ20PRL,Tyner21arXiv,
Nie22SB,Yuan21PRB,Wieder2020NC,Zeng2022FronPhys,Roy20PRR,LinMao18PRB,
ChenR21PRL,WangZM22arXiv,Wieder2020NC,Wang20PRL}. 3D quasicrystalline SOTSM phases can be achieved in a similar manner, i.e., by periodically staking 2D quasicrystalline SOTIs with an orbital-dependent hopping $t_z s_0\tau_3$ on each site~\cite{Fonseca2022arXiv}. After Fourier transformation applied to the vertical direction $z$, the 3D stacked Hamiltonian can be expressed as
\begin{equation}
H_{\text{3D}}=\sum_{k_z}H(M-2t_z\cos k_z).
\end{equation}
The conduction and valence bands of in this model have double degeneracy because of the presence of the combination of TRS and inversion symmetry $PT$~\cite{Roy20PRR,PRR2019Roy}, where $P=s_0\tau_3$ is the inversion symmetry operator. It is necessary to point out that when $\eta=2$, applying the stacked Hamiltonian to periodic cubic lattices will give birth to a 3D crystalline SOTSM~\cite{Roy20PRR,PRR2019Roy}~(see Appendix~\ref{Appendix1}) with four hinge Fermi arcs connecting the projection of fourfold degenerate Dirac points that are well defined in the momentum space. Next, we investigate the situation where the Hamiltonian is defined on a stack of 2D quasicrystals.

\begin{figure}[htp]
\centering
\includegraphics[width =\columnwidth]{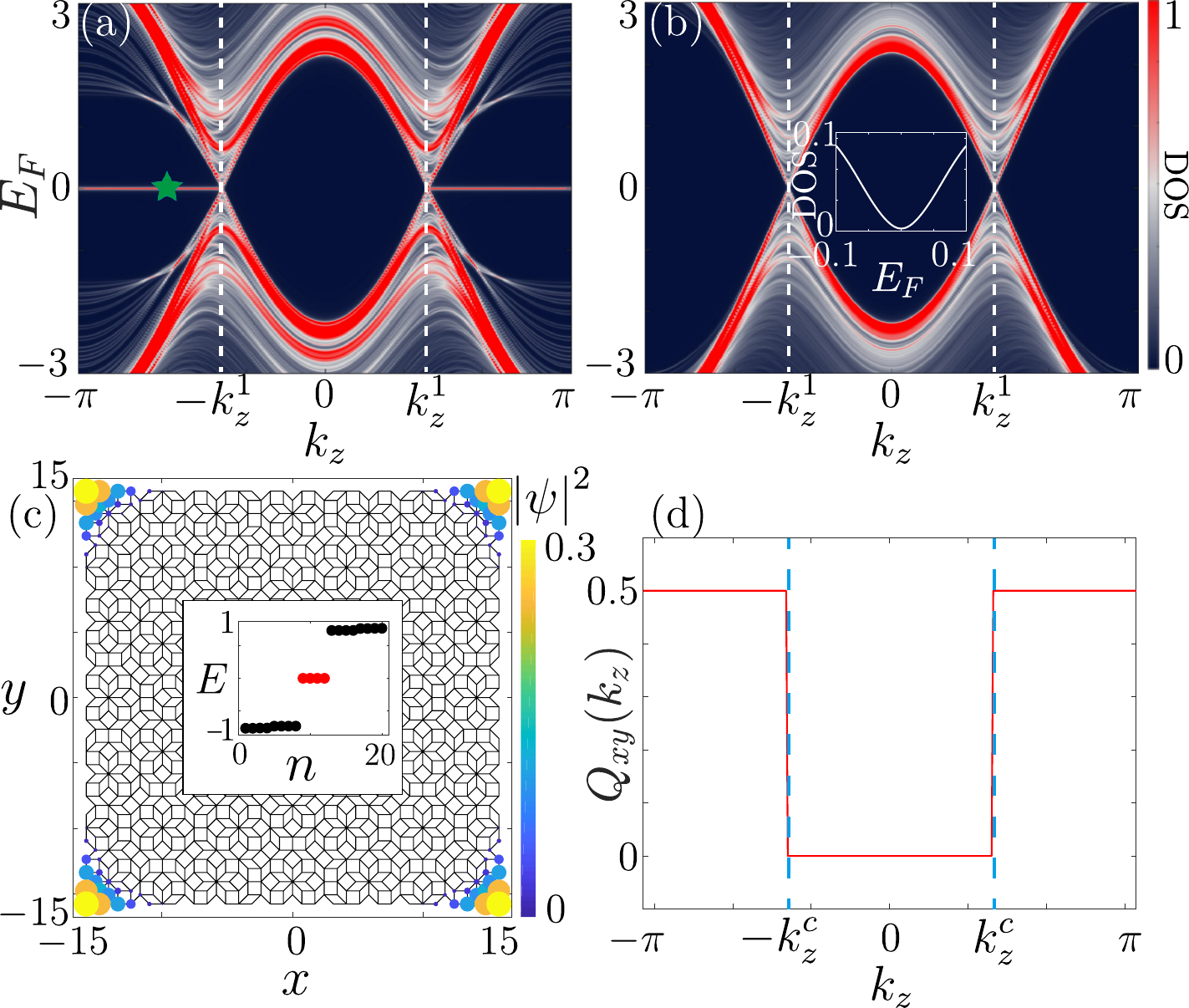}
\caption{Spectral function of the stacked AB-tiling quasicrystal with square-shaped disk [see Fig.~\ref{fig_illustration}(b)] as a function of $k_z$, under (a) open boundary condition in $xy$-plane and periodic boundary condition along the $z$-direction, and (b) periodic boundary conditions along all the three directions. (c) The probability distribution of the zero-energy modes with $k_z=-2$ [marked by the green star in (a)]. (d) The quadrupole moment as a function of $k_z$, calculated under periodic boundary conditions along all the three directions. The parameters are taken as $M=-2$, $t_2=1$, $g=1$, and $t_z=1.5$. The lattice site number is 1257.}
\label{fig_AB4}
\end{figure}

\subsection{$\eta=2$}

We first consider a 3D quasicrystal~[Fig.~\ref{fig_illustration}(b)] by stacking 2D AB-tiling quasicrystals with the square-shaped boundary~[Fig.~\ref{fig_illustration}(a)] and set the varying period of Wilson mass $\eta=2$. Figure~\ref{fig_AB4}(a) shows the spectral function $\mathcal{A} (E_F,k_z)$ of the 3D quasicrystalline system with open-boundary condition in the $xy$-plane. We can see that the bulk conduction and valence bands touch at two discrete points $k_z=\pm k_z^1$ where the energy gap is closed, indicating a semimetal phase. Importantly, fourfold degenerate zero-energy flat band boundary states emerge in the region $\left|k_z\right|>k_z^1$, describing hinge Fermi arc states in this semimetal phase. Figure~\ref{fig_AB4}(c) displays the probability density distribution of the zero-energy states at $k_z=-2$ [marked by the green star in Fig.~\ref{fig_AB4}(a)].

Figure~\ref{fig_AB4}(b) illustrates the spectral function of the quasicrystalline system with periodic boundary conditions along all the directions. The periodic boundary condition in the $xy$-plane is achieved by treating the system as a crystal with a supercell periodicity.  Comparing to the spectral function under open boundary condition in Fig.~\ref{fig_illustration}(a), the zero-energy flat band boundary states disappear, further confirming that the zero-energy modes in between $\pm k_z^1$ are hinge Fermi arc states.  Moreover, the higher-order topology of the hinge Fermi arcs is revealed by the quantized quadrupole moment $Q_{xy}=0.5$ for $\left|k_z\right|>k_z^1$ [Fig.~\ref{fig_AB4}(d)]. Therefore, the system is identified as a quasicrystalline SOTSM.

The bulk spectral function versus $k_z$ exhibits a linear dispersion near the gap closing points at $\pm k_z^1$ [Fig.~\ref{fig_AB4}(b)]. Meanwhile, the density of states around the gap closing points is parabolic, as shown in the inset of Fig.~\ref{fig_AB4}(b), which identifies the well-known bulk signatures of Dirac points in crystalline systems~\cite{Pixley2016PRX,Roy2018PRX}. These features suggest that the gapless points in the present system are Dirac points in quasicrystals. However, as discussed in Sec.~\ref{Sec_Gaplesspoint}, a more detailed analysis reveals that the situation is complex.

\begin{figure}[th]
\centering
\includegraphics[width =\columnwidth]{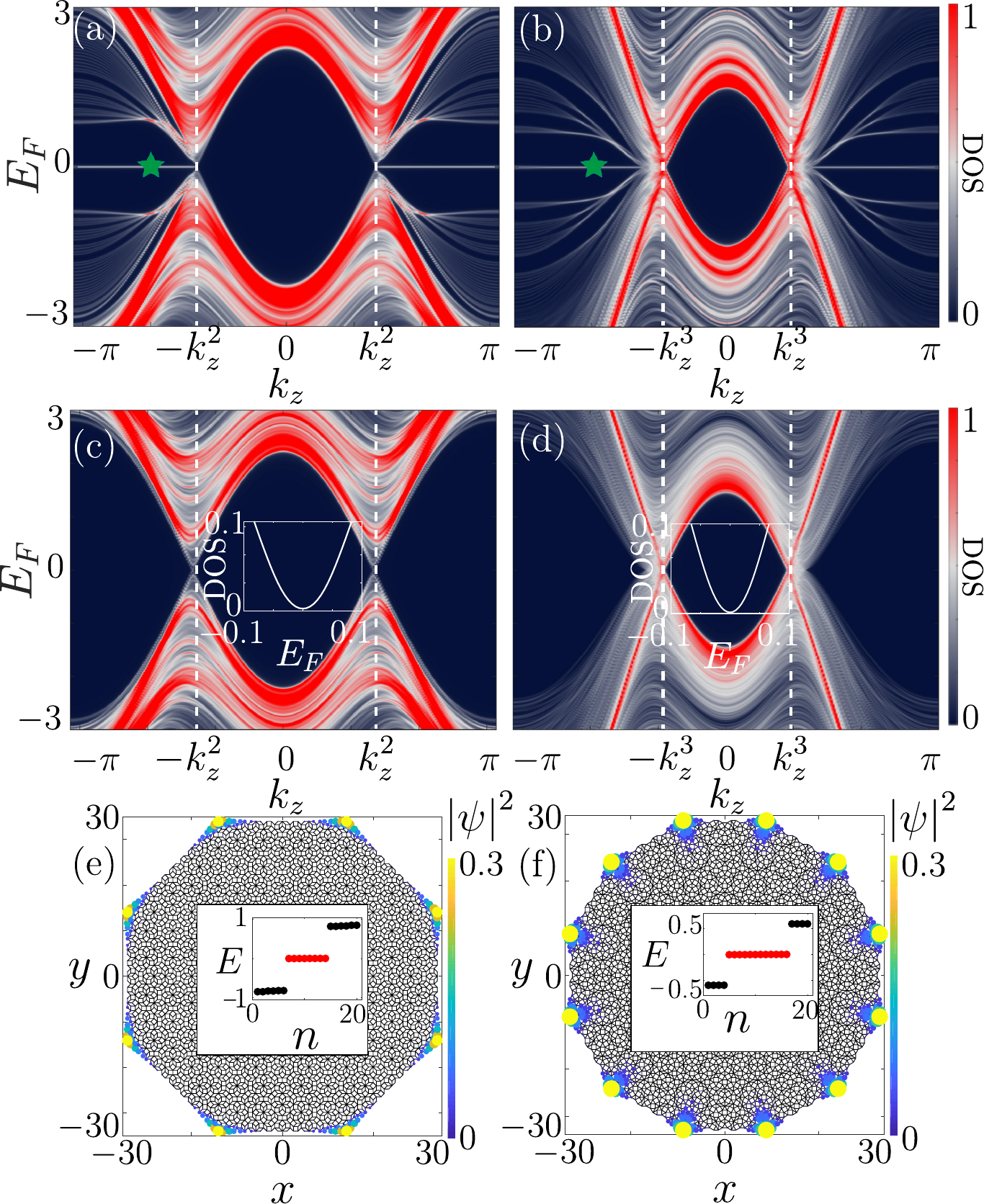}
\caption{Spectral function of the quasicrystalline octagonal system [see Fig.~\ref{fig_illustration}(d)] as a function of $k_z$ with (a) open boundary conditions along the in-plane $x$ and $y$ directions and periodic boundary condition along the $z$-direction, and (c) periodic boundary conditions along all the three directions. The lattice site number is 4713. (e) The probability distribution of the zero-energy modes in the quasicrystalline octagonal system with $k_z=-2$ [marked by the green star in (a)] . (b, d, f) are the same as (a, c, e), except that they describe the quasicrystalline dodecagonal system [see Fig.~\ref{fig_illustration}(f)] with the lattice site number is 4105. The parameters are $M=-2$, $t_2=1$, $g=1$, $t_z=1.5$ in (a, c, e), and $M=-3$, $t_2=2$, $g=2$, $t_z=2$ in (b, d, f).}
\label{fig_C8C12}
\end{figure}

\begin{figure*}[ht]
\centering
\includegraphics[width =1.7\columnwidth]{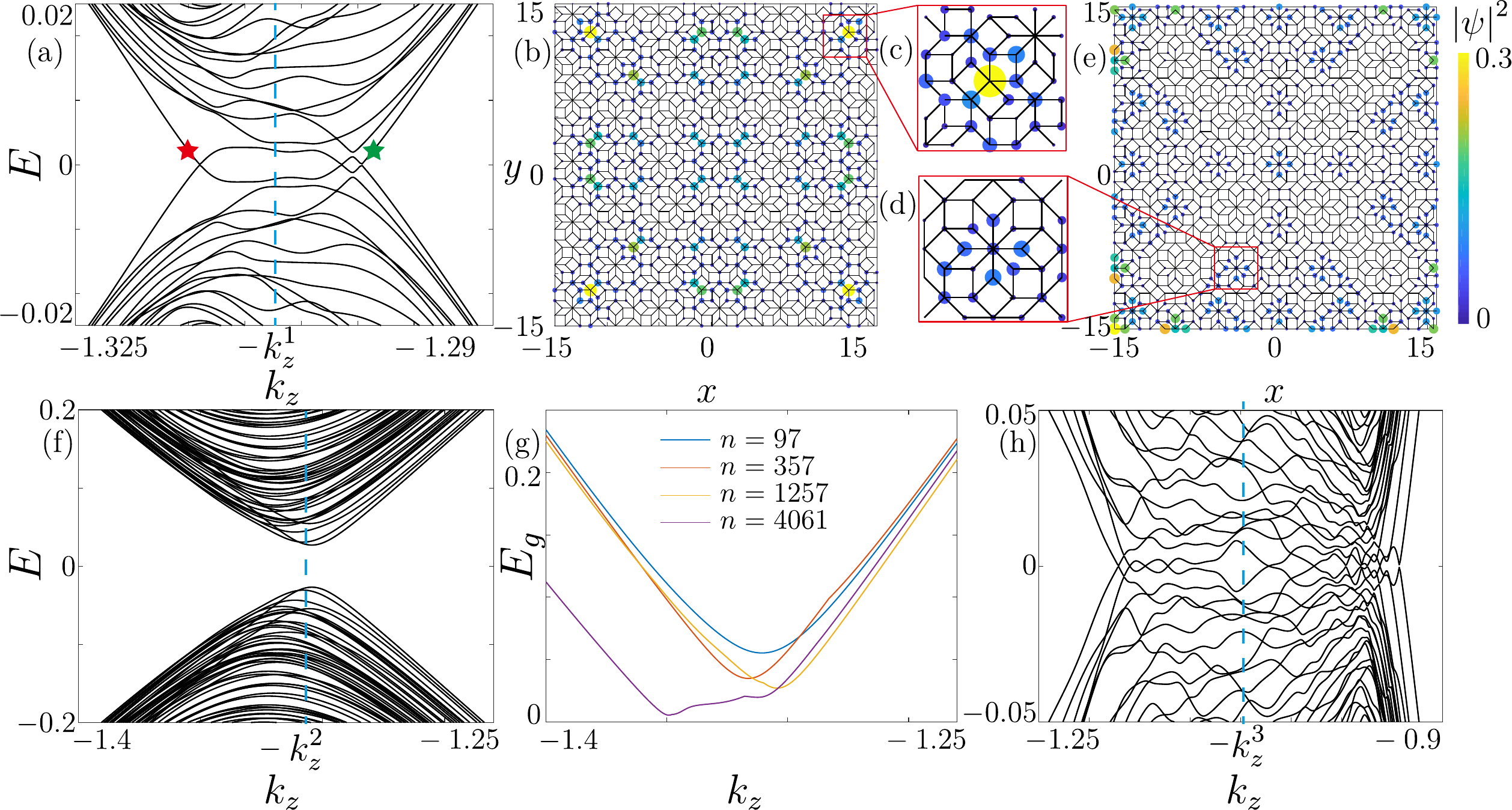}
\caption{(a) Energy spectrum versus $k_z$ of the quasicrystalline system near $-k_z^{1}$ with $\eta=2$ [see Fig.~\ref{fig_AB4}(b)].  (b) and (e) correspond to the probability distributions of the states marked by the red and green stars in (a), respectively. (c) and (d) show that the probability distributions in (b) and (e) are mainly distributed at two different kinds of local patches. (f) Spectrum of the quasicrystalline system for $k_z$ is near $-k_z^{2}$ [see Fig.~\ref{fig_C8C12}(c)] with $\eta=4$. (g) The energy gap $E_g$ as a function of $k_z$ for different system size with $\eta=4$. (h) Spectrum of the quasicrystalline system as a function of $k_z$ near $-k_z^{3}$ with $\eta=6$ [see Fig.~\ref{fig_C8C12}(d)]. }
\label{fig_Diracpoint}
\end{figure*}

\subsection{$\eta=4$ and $\eta=6$}
Now, we come to the case of $\eta=4$ and $\eta=6$, which can give rise to 2D quasicrystalline SOTIs without crystalline counterpart~\cite{ChenR20PRL,Hua20PRB}. Here, the 3D quasicrystalline systems are stacked by the AB-tiling octagonal quasicrystal [Figs.~\ref{fig_illustration}(c) and \ref{fig_illustration}(d)] and the Stampfli-tiling dodecagonal quasicrystal [Figs.~\ref{fig_illustration}(e) and \ref{fig_illustration}(f)], respectively. Figures~\ref{fig_C8C12}(a) and \ref{fig_C8C12}(b) show the spectral function $\mathcal{A} (E_F,k_z)$ of the two 3D quasicrystalline systems with open boundary condition in the $xy$-plane and periodic boundary condition along the vertical direction. The spectral functions look similar to that shown in Fig.~\ref{fig_AB4}(a), however, the degeneracy of zero-energy modes is different.
These zero-energy flat-band boundary modes in the region $\left|k_z\right|>k_z^1$ are hinge Fermi arc states traveling on the hinges of 3D octagonal/dodecagonal quasicrystals. This can be observed more clearly in Figs.~\ref{fig_C8C12}(e) and~\ref{fig_C8C12}(f), which show the energy spectra and the probability distributions of the zero-energy modes for fixed $k_z$ marked by the green stars shown in Figs.~\ref{fig_C8C12}(a) and \ref{fig_C8C12}(a), respectively. Apparently, the hinge Fermi arc states are inherited from the $C_n^z$-symmetric corner modes in quasicrystalline SOTIs, where $n=8$ in the AB-tiling octagonal quasicrystal and $n=12$ in the Stampfli-tiling dodecagonal quasicrystal.

To diagonalize the electronic structure of bulk state, we plot the spectral function under periodic boundary conditions along all the three directions in Figs.~\ref{fig_C8C12}(c) and \ref{fig_C8C12}(d). As seen in the case with $\eta=2$, similar phenomena are observed, such as the disappearance of zero-energy hinge arcs, a linear dispersion along $k_z$, and the quadratic density of states around the gap closing points.

Therefore, our study demonstrates that stacking 2D quasicrystals can result in the emergence of an exotic topological phase of matter i.e., the quasicrystalline SOTSMs, which possesses eight and twelve hinge Fermi arcs protected by forbidden rotation symmetries in crystalline systems. Our findings highlight the potential for stacking 2D quasicrystals and expand our understanding of condensed matter physics.

\section{Dirac-like points}
\label{Sec_Gaplesspoint}
Upon initial inspection, the gap closing points near $k_z=\pm k_z^{1,2,3}$ shown in Figs.~\ref{fig_AB4}(b), \ref{fig_C8C12}(c) and \ref{fig_C8C12}(d) are reminiscent of the Dirac point characterized by the massless Dirac equation. They both exhibit a linear dispersion along $k_z$ and a unique quadratic density of state near the gap closing points. However, a closer inspection of the spectrum reveals that the gap closing points in quasicrystalline SOTSMs are distinct from those in crystalline second-order topological Dirac semimetals (SODSMs).

Figure~\ref{fig_Diracpoint}(a) shows the spectrum near the gap closing point $k_z^{1}$ in the SOTSM of $\eta=2$ under periodic boundary conditions along all the directions [see Fig.~\ref{fig_AB4}(b)]. There appear three band-crossing points, which is quite different from the crystalline SODSM phase that hosts only one band-crossing point~[Fig.~\ref{fig_square}(e)]. Figures~\ref{fig_Diracpoint}(b) and \ref{fig_Diracpoint}(e) show the wave function of the states marked by the red and green stars in Fig.~\ref{fig_Diracpoint}(a), respectively. One of the band-crossing is dominated by the local patch containing three square tiles and two rhombus tiles [Figs.~\ref{fig_Diracpoint}(b) and~\ref{fig_Diracpoint}(c)], and the other band-crossing is dominated by the local patch containing six rhombus tiles [Figs.~\ref{fig_Diracpoint}(d) and~\ref{fig_Diracpoint}(e)]. The appearance of multiple band-crossing points is because gap closes at different $k_z$ for distinct kinds of local patches. This phenomenon is attributed to the absence of discrete translational symmetry in quasicrystalline systems.

For the AB-tiling octagonal quasicrystal with $\eta=4$, the spectrum opens a tiny energy gap~[Fig.~\ref{fig_Diracpoint}(f)]. The size of the energy gap decays with the increase of size~[Fig.~\ref{fig_Diracpoint}(g)]. For the Stampfli-tiling quasicrystal with $\eta=6$, the spectrum is similar to the case with $\eta=2$, except that there appear more band crossings. This is because there are more different patterns of local patches in Stampfli-tiling quasicrystal.

Though the gap-closing points in quasicrystalline SOTSMs manifest several similarities compared to the Dirac points in crystalline SODSMs. However, we found the fine structure of the gap-closing points due to the absence of translational symmetry by further checking the spectrum. Therefore, we dub these gap closing points in the quasicrystalline SOTSM phase as Dirac-like points.

\section{Phase diagram}
\label{Phase}
\begin{figure}[t]
\centering
\includegraphics[width =\columnwidth]{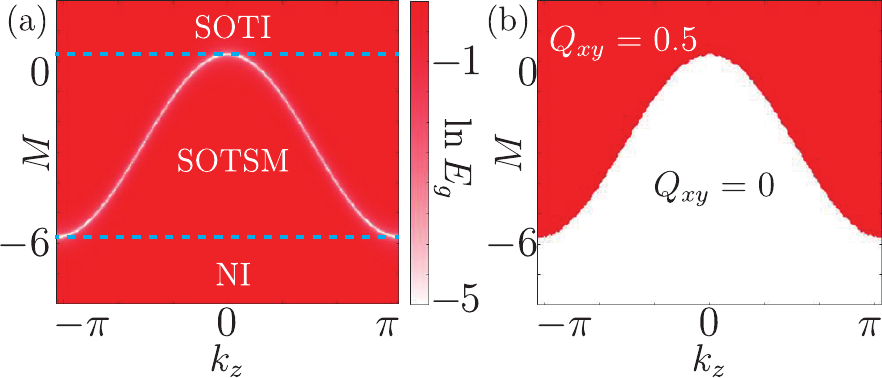}
\caption{(a) The logarithm of the energy gap $\ln{E_g}$ and (b) the quadrupole moment $Q_{xy}$ as functions of the parameter $M$ and the momentum $k_z$. In (a), each point in the white line depicts the gap-closing point. In (b), the red and white areas host a quantized quadrupole moment $Q_{xy}=0.5$ and a zero quadrupole moment $Q_{xy}=0$, respectively. Depending on $M$, the system can be divided into three phases: the SOTI phase, the SOTSM phase, and the normal insulator (NI) phase. The three phases are separated by the dashed cyan lines. The results are obtained in the AB-tiling quasicrystalline square systems with periodic boundary conditions along all the directions. The lattice site number is 1257. The parameters are $\eta=2$, $t_2=1$, $g=1$, and $t_z=1.5$.  }
\label{fig_phase}
\end{figure}

\begin{figure}[th!]
\centering
\includegraphics[width =1\columnwidth]{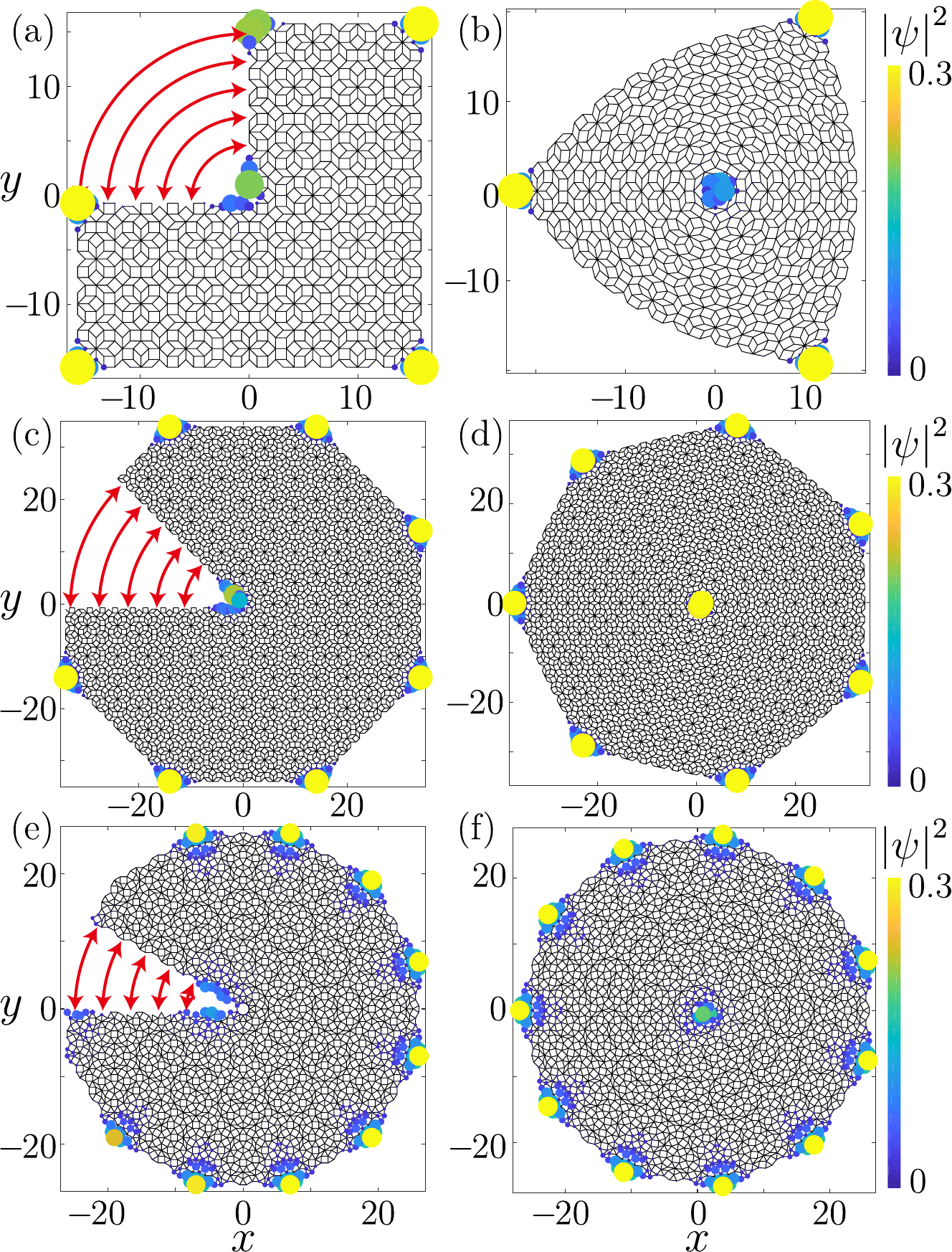}
\caption{Probability distributions of the zero-energy modes in the different systems with $k_z=-2$. (a, c, e) Systems with a certain segment being cut out. (b, d, f) Disclination systems where the two sides of the cut is glued together [the red lines in (a, c, e)].}
\label{fig_defect}
\end{figure}
We present the topological phase diagram of the stacked quasicrystal system in this section.
Figures~\ref{fig_phase}(a)-\ref{fig_phase}(b) show $\ln E_g$ and $Q_{xy}$ as functions of the momentum $k_z$ and the parameter $M$ for the AB-tiling quasicrystalline square system with $\eta=2$. $E_g$ is the value of the energy gap obtained under periodic boundary conditions along all the three directions. Each point along the white line corresponds to the gap-closing point shown in Fig.~\ref{fig_AB4}(b). For about $-5.7<M<0.3$, the existence of the gap closure with the accompanying topological phase transition between $Q_{xy}=0$ and $Q_{xy}=0.5$ indicates that the system corresponds to the SOTSM phase.
For about $M>0.3$, the system corresponds to a 3D quasicrystalline SOTI phase with a topological gap characterized by a quantized quadrupole moment $Q_{xy}=0.5$ for any $k_z$. For about $M<-5.7$, the system is a normal insulator (NI) with a topologically trivial gap.

Above we only consider the case of $\eta=2$ in the AB-tiling quasicrystal. In the cases of the AB-tiling octagonal quasicrystal with $\eta=4$ and the Stampfli-tiling dodecagonal quasicrystal with $\eta=6$, we find similar results by adjusting the parameter $M$, i.e., the systems also support the quasicrystalline SOTSM phase, the 3D quasicrystalline SOTI phase, and the NI phase.

\section{Disclination-induced bound states}
\label{Defect}

Disclination-induced bound states provide a potential probe of crystalline topology, which has been widely investigated in different topological systems~\cite{Benalcazar19PRB,Miert18PRB,Varjas19PRL,Geier2021SPP,Qi2022APL}. Recently, disclination-induced bound states have been observed in topological crystalline insulators~\cite{Peterson2021Nature}, acoustic topological insulators~\cite{Ye2021NC,Xue21PRL,Deng22PRL,Xia2022AMS}, and acoustic Weyl semimetals~\cite{WangQ2021NC}. In this section, we study the disclination-induced bound states in the quasicrystalline SOTSM phase.

The disclination is introduced by cutting out a specific segment [the first column in Fig.~\ref{fig_defect}] and then glue the lattice back together [the second column in Fig.~\ref{fig_defect}]. The two sides of the cut are glued together by identifying sites on the two sides of the cut related by rotational symmetry, which is called a Volterra process~\cite{Varjas19PRL,Geier2021SPP,Kleman08RMP}. The defects break the rotational symmetry locally at the center of lattice, but the rest preserves the rotational symmetry and is indistinguishable from the bulk of the original system without the cut.

The corresponding spectral function of sample geometries in Figs.~\ref{fig_defect}(a)-\ref{fig_defect}(b), Figs.~\ref{fig_defect}(c)-\ref{fig_defect}(d), and Figs.~\ref{fig_defect}(e)-\ref{fig_defect}(f) are similar to Fig.~\ref{fig_AB4}(a), Fig.~\ref{fig_C8C12}(a), and Fig.~\ref{fig_C8C12}(b), respectively, except that the spatial probability distributions are different for the zero-energy modes. The colored points in Fig.~\ref{fig_defect} display the probability distributions of the zero-energy modes in these systems with $k_z=-2$.
For the three different disclination systems in Figs.~\ref{fig_defect}(b), \ref{fig_defect}(d), and \ref{fig_defect}(f), they both host one zero-energy mode at the disclination core, and three, seven, and eleven zero-energy modes at the hinges of the systems, respectively. Moreover, similar to the zero-energy hinge modes, the disclination modes only appear for $\left|k_z\right|>k_z^{1/2/3}$, and disappear in the regions of $\left|k_z\right|<k_z^{1/2/3}$. This further reveals that the disclination-induced bound states and the hinge Fermi arc states are the consequence of nontrivial bulk topology, which cannot be removed without topologically trivializing the bulk of systems~\cite{Varjas19PRL}. Moreover, the $k_z$-dependent disclination-induced bound states provide an experimental probe for the quasicrystalline SOTSM phase.

\section{Conclusion and discussion}
\label{discussion}

In conclusion, this study has demonstrated that a stack of 2D quasicrystalline SOTIs can give rise to 3D quasicrystalline SOTSM phases. These 3D phases exhibit rotation-symmetry protected hinge Fermi arcs, which are forbidden in crystalline systems. Additionally, our calculations have shown that the stacked systems also support the 3D quasicrystalline SOTI phase, as evidenced by the phase diagram. We have proposed that the dependence of $k_z$ on disclination-induced bound states can serve as an experimental probe for the quasicrystalline SOTSM phase.

While the quasicrystalline SOTSM shares similarities with the crystalline SODSM~\cite{Roy20PRR,LinMao18PRB,Yuan21PRB,Zeng2022FronPhys,Wieder2020NC,Wang20PRL}, there are three main distinctions between them. Firstly, the number of $C_n^z$-symmetry protected hinge Fermi arcs in the quasicrystalline SOTSM  is not limited to four, as observed in crystalline SODSM, but can be eight or twelve as well. Secondly, in the quasicrystalline SOTSM, the lack of translational symmetry renders the in-plane momentum ineffective as a quantum number, making it impossible to define Dirac points in momentum space, unlike in crystalline SODSM where the Dirac equation applies. Lastly, the spectrum of the quasicrystalline SOTSM exhibits a higher number of band-crossing points compared to the crystalline SODSM, a consequence of the absence of in-plane translational symmetry in the stacked quasicrystals.

Moreover, recent experiments investigating the stack of Ta$_{1.6}$Te quasicrystal layers~\cite{Cain2020PNAS}, along with first-principles calculations and symmetry analysis, have revealed a symmetry-protected semimetal phase and explored the topological properties of the material. This suggests that the quasicrystalline SOTSM phase can be experimentally realized in real materials. Furthermore, considering the successful experimental realization of the 2D quasicrystalline SOTI phase in electrical circuit systems~\cite{Lv2021ComPhys}, we believe that the quasicrystalline SOTSM holds promise in metamaterials. These unique features and possibilities offer exciting prospects for the future implementation of our proposal.

\begin{acknowledgments}
D.-H.X. was supported by the NSFC (under Grant Nos.~12074108 and 12147102), the Natural Science Foundation of Chongqing (Grant No.~CSTB2022NSCQ-MSX0568). R.C. acknowledges the support of the Chutian Scholars Program in Hubei Province.  B.Z. was supported by the NSFC (under Grant No. 12074107), the program of outstanding young and middle-aged scientific and technological innovation team of colleges and universities in Hubei Province (under Grant No. T2020001) and the innovation group project of the natural science foundation of Hubei Province of China (under Grant No. 2022CFA012).
\end{acknowledgments}

\appendix
\section{Crystalline SODSM}\label{Appendix1}

\begin{figure}
\centering
\includegraphics[width =\columnwidth]{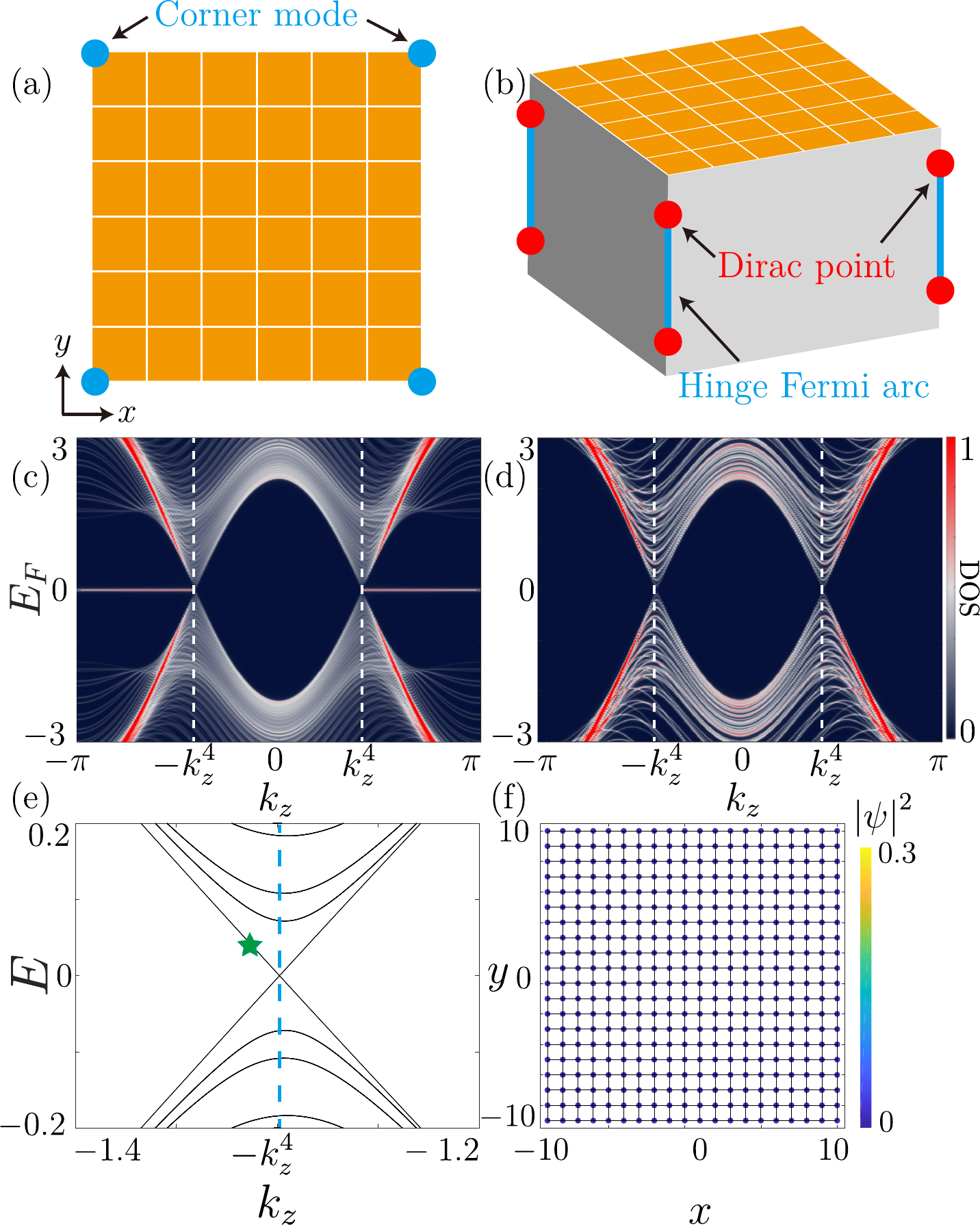}
\caption{(a) Schematic illustrations of the crystalline SOTI in a square lattice. The 2D crystalline system supports four rotation-symmetry-protected zero-energy corner modes (the cyan points). (b) A simple stack of the 2D crystalline system can lead to the 3D crystalline SODSM phase with four rotation-symmetry-protected hinge Fermi arc states (the cyan lines). (c-d) Spectral function of the crystalline square system as a function of $k_z$ with (c) open boundary conditions along the in-plane directions and periodic boundary condition along the $z$-direction and (d) periodic boundary conditions along all the directions. (e) Energy spectrum of the quasicrystalline system for $k_z$ is near $-k_z^{4}$. (f) Probability distribution of the states marked by the green star in (e).
The parameters are taken as $\eta=2$, $M=-2$, $t_2=1$, $g=1$, and $t_z=1.5$. The lattice number is 400.}
\label{fig_square}

\end{figure}

To make a comparative study, we investigate the 3D crystalline SODSM phase~[Fig.~\ref{fig_square}(b)], modeled by staking 2D crystalline SOTIs along the vertical direction~[Fig.~\ref{fig_square}(a)]. Figures~\ref{fig_square}(c) and \ref{fig_square}(d) show the spectral function of the crystalline system with open and periodic boundary conditions in $xy$-plane, respectively. Hinge Fermi arcs appear and connect the band-closing points at $k_z=\pm k_z^4$. The results are similar to those in Figs.~\ref{fig_AB4}(a)-\ref{fig_AB4}(b). Figure~\ref{fig_square}(e) shows the spectrum near the band-closing point $-k_c^{4}$. Only one band-crossing point is observed because of the existence of transitional symmetry in crystalline systems. This is observed more clearly in Fig.~\ref{fig_square}(f). The
probability distribution of the state labeled by green star [Fig.~\ref{fig_square}(e)] is uniformly distributed and all the local patches undergo the topological phase transition simultaneously when $k_z$ varies.
Moreover, we find that the low-energy effective Hamiltonian can be described by the massless Dirac equation. Therefore, the system is identified as the crystalline SODSM phase.

\bibliographystyle{apsrev4-1-etal-title_6authors}
\bibliography{../refs-transport}

\begin{thebibliography}{80}%
\makeatletter
\providecommand \@ifxundefined [1]{%
 \@ifx{#1\undefined}
}%
\providecommand \@ifnum [1]{%
 \ifnum #1\expandafter \@firstoftwo
 \else \expandafter \@secondoftwo
 \fi
}%
\providecommand \@ifx [1]{%
 \ifx #1\expandafter \@firstoftwo
 \else \expandafter \@secondoftwo
 \fi
}%
\providecommand \natexlab [1]{#1}%
\providecommand \enquote  [1]{``#1''}%
\providecommand \bibnamefont  [1]{#1}%
\providecommand \bibfnamefont [1]{#1}%
\providecommand \citenamefont [1]{#1}%
\providecommand \href@noop [0]{\@secondoftwo}%
\providecommand \href [0]{\begingroup \@sanitize@url \@href}%
\providecommand \@href[1]{\@@startlink{#1}\@@href}%
\providecommand \@@href[1]{\endgroup#1\@@endlink}%
\providecommand \@sanitize@url [0]{\catcode `\\12\catcode `\$12\catcode
  `\&12\catcode `\#12\catcode `\^12\catcode `\_12\catcode `\%12\relax}%
\providecommand \@@startlink[1]{}%
\providecommand \@@endlink[0]{}%
\providecommand \url  [0]{\begingroup\@sanitize@url \@url }%
\providecommand \@url [1]{\endgroup\@href {#1}{\urlprefix }}%
\providecommand \urlprefix  [0]{URL }%
\providecommand \Eprint [0]{\href }%
\providecommand \doibase [0]{http://dx.doi.org/}%
\providecommand \selectlanguage [0]{\@gobble}%
\providecommand \bibinfo  [0]{\@secondoftwo}%
\providecommand \bibfield  [0]{\@secondoftwo}%
\providecommand \translation [1]{[#1]}%
\providecommand \BibitemOpen [0]{}%
\providecommand \bibitemStop [0]{}%
\providecommand \bibitemNoStop [0]{.\EOS\space}%
\providecommand \EOS [0]{\spacefactor3000\relax}%
\providecommand \BibitemShut  [1]{\csname bibitem#1\endcsname}%
\let\auto@bib@innerbib\@empty
\bibitem [{\citenamefont {Benalcazar}\ \emph
  {et~al.}(2017{\natexlab{a}})\citenamefont {Benalcazar}, \citenamefont
  {Bernevig},\ and\ \citenamefont {Hughes}}]{Benalcazar2017Science}%
  \BibitemOpen
  \bibfield  {author} {\bibinfo {author} {\bibfnamefont {W.~A.}\ \bibnamefont
  {Benalcazar}}, \bibinfo {author} {\bibfnamefont {B.~A.}\ \bibnamefont
  {Bernevig}}, \ and\ \bibinfo {author} {\bibfnamefont {T.~L.}\ \bibnamefont
  {Hughes}},\ }\bibfield  {title} {\enquote {\bibinfo {title} {Quantized
  electric multipole insulators}}, }\href {\doibase 10.1126/science.aah6442}
  {\bibfield  {journal} {\bibinfo  {journal} {Science}\ }\textbf {\bibinfo
  {volume} {357}},\ \bibinfo {pages} {61} (\bibinfo {year}
  {2017}{\natexlab{a}})}\BibitemShut {NoStop}%
\bibitem [{\citenamefont {Schindler}\ \emph {et~al.}(2018)\citenamefont
  {Schindler}, \citenamefont {Cook}, \citenamefont {Vergniory}, \citenamefont
  {Wang}, \citenamefont {Parkin}, \citenamefont {Bernevig},\ and\ \citenamefont
  {Neupert}}]{Schindler2018SciAdv}%
  \BibitemOpen
  \bibfield  {author} {\bibinfo {author} {\bibfnamefont {F.}~\bibnamefont
  {Schindler}}, \bibinfo {author} {\bibfnamefont {A.~M.}\ \bibnamefont {Cook}},
  \bibinfo {author} {\bibfnamefont {M.~G.}\ \bibnamefont {Vergniory}}, \bibinfo
  {author} {\bibfnamefont {Z.}~\bibnamefont {Wang}}, \bibinfo {author}
  {\bibfnamefont {S.~S.~P.}\ \bibnamefont {Parkin}}, \bibinfo {author}
  {\bibfnamefont {B.~A.}\ \bibnamefont {Bernevig}}, \ and\ \bibinfo {author}
  {\bibfnamefont {T.}~\bibnamefont {Neupert}},\ }\bibfield  {title} {\enquote
  {\bibinfo {title} {Higher-order topological insulators}}, }\href {\doibase
  10.1126/sciadv.aat0346} {\bibfield  {journal} {\bibinfo  {journal} {Sci.
  Adv.}\ }\textbf {\bibinfo {volume} {4}},\ \bibinfo {pages} {eaat0346}
  (\bibinfo {year} {2018})}\BibitemShut {NoStop}%
\bibitem [{\citenamefont {Langbehn}\ \emph {et~al.}(2017)\citenamefont
  {Langbehn}, \citenamefont {Peng}, \citenamefont {Trifunovic}, \citenamefont
  {von Oppen},\ and\ \citenamefont {Brouwer}}]{Langbehn17PRL}%
  \BibitemOpen
  \bibfield  {author} {\bibinfo {author} {\bibfnamefont {J.}~\bibnamefont
  {Langbehn}}, \bibinfo {author} {\bibfnamefont {Y.}~\bibnamefont {Peng}},
  \bibinfo {author} {\bibfnamefont {L.}~\bibnamefont {Trifunovic}}, \bibinfo
  {author} {\bibfnamefont {F.}~\bibnamefont {von Oppen}}, \ and\ \bibinfo
  {author} {\bibfnamefont {P.~W.}\ \bibnamefont {Brouwer}},\ }\bibfield
  {title} {\enquote {\bibinfo {title} {Reflection-symmetric second-order
  topological insulators and superconductors}}, }\href {\doibase
  10.1103/PhysRevLett.119.246401} {\bibfield  {journal} {\bibinfo  {journal}
  {Phys. Rev. Lett.}\ }\textbf {\bibinfo {volume} {119}},\ \bibinfo {pages}
  {246401} (\bibinfo {year} {2017})}\BibitemShut {NoStop}%
\bibitem [{\citenamefont {Benalcazar}\ \emph
  {et~al.}(2017{\natexlab{b}})\citenamefont {Benalcazar}, \citenamefont
  {Bernevig},\ and\ \citenamefont {Hughes}}]{Benalcazar17PRB}%
  \BibitemOpen
  \bibfield  {author} {\bibinfo {author} {\bibfnamefont {W.~A.}\ \bibnamefont
  {Benalcazar}}, \bibinfo {author} {\bibfnamefont {B.~A.}\ \bibnamefont
  {Bernevig}}, \ and\ \bibinfo {author} {\bibfnamefont {T.~L.}\ \bibnamefont
  {Hughes}},\ }\bibfield  {title} {\enquote {\bibinfo {title} {Electric
  multipole moments, topological multipole moment pumping, and chiral hinge
  states in crystalline insulators}}, }\href {\doibase
  10.1103/PhysRevB.96.245115} {\bibfield  {journal} {\bibinfo  {journal} {Phys.
  Rev. B}\ }\textbf {\bibinfo {volume} {96}},\ \bibinfo {pages} {245115}
  (\bibinfo {year} {2017}{\natexlab{b}})}\BibitemShut {NoStop}%
\bibitem [{\citenamefont {Song}\ \emph {et~al.}(2017)\citenamefont {Song},
  \citenamefont {Fang},\ and\ \citenamefont {Fang}}]{Song17PRL}%
  \BibitemOpen
  \bibfield  {author} {\bibinfo {author} {\bibfnamefont {Z.}~\bibnamefont
  {Song}}, \bibinfo {author} {\bibfnamefont {Z.}~\bibnamefont {Fang}}, \ and\
  \bibinfo {author} {\bibfnamefont {C.}~\bibnamefont {Fang}},\ }\bibfield
  {title} {\enquote {\bibinfo {title} {$(d\ensuremath{-}2)$-dimensional edge
  states of rotation symmetry protected topological states}}, }\href {\doibase
  10.1103/PhysRevLett.119.246402} {\bibfield  {journal} {\bibinfo  {journal}
  {Phys. Rev. Lett.}\ }\textbf {\bibinfo {volume} {119}},\ \bibinfo {pages}
  {246402} (\bibinfo {year} {2017})}\BibitemShut {NoStop}%
\bibitem [{\citenamefont {Li}\ \emph {et~al.}(2020)\citenamefont {Li},
  \citenamefont {Wang}, \citenamefont {Li}, \citenamefont {Zheng},
  \citenamefont {Brinkman}, \citenamefont {Yu},\ and\ \citenamefont
  {Liao}}]{LiCZ20PRL}%
  \BibitemOpen
  \bibfield  {author} {\bibinfo {author} {\bibfnamefont {C.-Z.}\ \bibnamefont
  {Li}}, \bibinfo {author} {\bibfnamefont {A.-Q.}\ \bibnamefont {Wang}},
  \bibinfo {author} {\bibfnamefont {C.}~\bibnamefont {Li}}, \bibinfo {author}
  {\bibfnamefont {W.-Z.}\ \bibnamefont {Zheng}}, \bibinfo {author}
  {\bibfnamefont {A.}~\bibnamefont {Brinkman}}, \bibinfo {author}
  {\bibfnamefont {D.-P.}\ \bibnamefont {Yu}}, \ and\ \bibinfo {author}
  {\bibfnamefont {Z.-M.}\ \bibnamefont {Liao}},\ }\bibfield  {title} {\enquote
  {\bibinfo {title} {Reducing electronic transport dimension to topological
  hinge states by increasing geometry size of {Dirac} semimetal {Josephson}
  junctions}}, }\href {\doibase 10.1103/PhysRevLett.124.156601} {\bibfield
  {journal} {\bibinfo  {journal} {Phys. Rev. Lett.}\ }\textbf {\bibinfo
  {volume} {124}},\ \bibinfo {pages} {156601} (\bibinfo {year}
  {2020})}\BibitemShut {NoStop}%
\bibitem [{\citenamefont {Tyner}\ \emph {et~al.}(2021)\citenamefont {Tyner},
  \citenamefont {Sur}, \citenamefont {Zhou}, \citenamefont {Puggioni},
  \citenamefont {Darancet}, \citenamefont {Rondinelli},\ and\ \citenamefont
  {Goswami}}]{Tyner21arXiv}%
  \BibitemOpen
  \bibfield  {author} {\bibinfo {author} {\bibfnamefont {A.~C.}\ \bibnamefont
  {Tyner}}, \bibinfo {author} {\bibfnamefont {S.}~\bibnamefont {Sur}}, \bibinfo
  {author} {\bibfnamefont {Q.}~\bibnamefont {Zhou}}, \bibinfo {author}
  {\bibfnamefont {D.}~\bibnamefont {Puggioni}}, \bibinfo {author}
  {\bibfnamefont {P.}~\bibnamefont {Darancet}}, \bibinfo {author}
  {\bibfnamefont {J.~M.}\ \bibnamefont {Rondinelli}}, \ and\ \bibinfo {author}
  {\bibfnamefont {P.}~\bibnamefont {Goswami}},\ }\bibfield  {title} {\enquote
  {\bibinfo {title} {{Non-Abelian Stokes theorem and quantized Berry flux}}},
  }\href {https://arxiv.org/abs/2102.06207} {\bibfield  {journal} {\bibinfo
  {journal} {arXiv:2102.06207}\ } (\bibinfo {year} {2021})}\BibitemShut
  {NoStop}%
\bibitem [{\citenamefont {Nie}\ \emph {et~al.}(2022)\citenamefont {Nie},
  \citenamefont {Chen}, \citenamefont {Yue}, \citenamefont {Le}, \citenamefont
  {Yuan}, \citenamefont {Wang}, \citenamefont {Zhang},\ and\ \citenamefont
  {Weng}}]{Nie22SB}%
  \BibitemOpen
  \bibfield  {author} {\bibinfo {author} {\bibfnamefont {S.}~\bibnamefont
  {Nie}}, \bibinfo {author} {\bibfnamefont {J.}~\bibnamefont {Chen}}, \bibinfo
  {author} {\bibfnamefont {C.}~\bibnamefont {Yue}}, \bibinfo {author}
  {\bibfnamefont {C.}~\bibnamefont {Le}}, \bibinfo {author} {\bibfnamefont
  {D.}~\bibnamefont {Yuan}}, \bibinfo {author} {\bibfnamefont {Z.}~\bibnamefont
  {Wang}}, \bibinfo {author} {\bibfnamefont {W.}~\bibnamefont {Zhang}}, \ and\
  \bibinfo {author} {\bibfnamefont {H.}~\bibnamefont {Weng}},\ }\bibfield
  {title} {\enquote {\bibinfo {title} {{Tunable Dirac semimetals with
  higher-order Fermi arcs in Kagome lattices {Pd$_3$Pb$_2$X$_2$ (X=S,Se)}}}},
  }\href {\doibase https://doi.org/10.1016/j.scib.2022.09.003} {\bibfield
  {journal} {\bibinfo  {journal} {Science Bulletin}\ }\textbf {\bibinfo
  {volume} {67}},\ \bibinfo {pages} {1958} (\bibinfo {year}
  {2022})}\BibitemShut {NoStop}%
\bibitem [{\citenamefont {Fang}\ and\ \citenamefont {Cano}(2021)}]{Yuan21PRB}%
  \BibitemOpen
  \bibfield  {author} {\bibinfo {author} {\bibfnamefont {Y.}~\bibnamefont
  {Fang}}\ and\ \bibinfo {author} {\bibfnamefont {J.}~\bibnamefont {Cano}},\
  }\bibfield  {title} {\enquote {\bibinfo {title} {{Classification of Dirac
  points with higher-order Fermi arcs}}}, }\href {\doibase
  10.1103/PhysRevB.104.245101} {\bibfield  {journal} {\bibinfo  {journal}
  {Phys. Rev. B}\ }\textbf {\bibinfo {volume} {104}},\ \bibinfo {pages}
  {245101} (\bibinfo {year} {2021})}\BibitemShut {NoStop}%
\bibitem [{\citenamefont {Wieder}\ \emph {et~al.}(2020)\citenamefont {Wieder},
  \citenamefont {Wang}, \citenamefont {Cano}, \citenamefont {Dai},
  \citenamefont {Schoop}, \citenamefont {Bradlyn},\ and\ \citenamefont
  {Bernevig}}]{Wieder2020NC}%
  \BibitemOpen
  \bibfield  {author} {\bibinfo {author} {\bibfnamefont {B.~J.}\ \bibnamefont
  {Wieder}}, \bibinfo {author} {\bibfnamefont {Z.}~\bibnamefont {Wang}},
  \bibinfo {author} {\bibfnamefont {J.}~\bibnamefont {Cano}}, \bibinfo {author}
  {\bibfnamefont {X.}~\bibnamefont {Dai}}, \bibinfo {author} {\bibfnamefont
  {L.~M.}\ \bibnamefont {Schoop}}, \bibinfo {author} {\bibfnamefont
  {B.}~\bibnamefont {Bradlyn}}, \ and\ \bibinfo {author} {\bibfnamefont
  {B.~A.}\ \bibnamefont {Bernevig}},\ }\bibfield  {title} {\enquote {\bibinfo
  {title} {{Strong and fragile topological Dirac semimetals with higher-order
  Fermi arcs}}}, }\href {\doibase 10.1038/s41467-020-14443-5} {\bibfield
  {journal} {\bibinfo  {journal} {Nat. Commun.}\ }\textbf {\bibinfo {volume}
  {11}},\ \bibinfo {pages} {627} (\bibinfo {year} {2020})}\BibitemShut
  {NoStop}%
\bibitem [{\citenamefont {Zeng}\ \emph {et~al.}(2023)\citenamefont {Zeng},
  \citenamefont {Chen}, \citenamefont {Chen}, \citenamefont {Liu},
  \citenamefont {Sheng},\ and\ \citenamefont {Yang}}]{Zeng2022FronPhys}%
  \BibitemOpen
  \bibfield  {author} {\bibinfo {author} {\bibfnamefont {X.-T.}\ \bibnamefont
  {Zeng}}, \bibinfo {author} {\bibfnamefont {Z.}~\bibnamefont {Chen}}, \bibinfo
  {author} {\bibfnamefont {C.}~\bibnamefont {Chen}}, \bibinfo {author}
  {\bibfnamefont {B.-B.}\ \bibnamefont {Liu}}, \bibinfo {author} {\bibfnamefont
  {X.-L.}\ \bibnamefont {Sheng}}, \ and\ \bibinfo {author} {\bibfnamefont
  {S.~A.}\ \bibnamefont {Yang}},\ }\bibfield  {title} {\enquote {\bibinfo
  {title} {{Topological hinge modes in Dirac semimetals}}}, }\href {\doibase
  10.1007/s11467-022-1221-y} {\bibfield  {journal} {\bibinfo  {journal} {Front.
  Phys.}\ }\textbf {\bibinfo {volume} {18}},\ \bibinfo {pages} {13308}
  (\bibinfo {year} {2023})}\BibitemShut {NoStop}%
\bibitem [{\citenamefont {Szab\'o}\ and\ \citenamefont {Roy}(2020)}]{Roy20PRR}%
  \BibitemOpen
  \bibfield  {author} {\bibinfo {author} {\bibfnamefont {A.~L.}\ \bibnamefont
  {Szab\'o}}\ and\ \bibinfo {author} {\bibfnamefont {B.}~\bibnamefont {Roy}},\
  }\bibfield  {title} {\enquote {\bibinfo {title} {Dirty higher-order dirac
  semimetal: Quantum criticality and bulk-boundary correspondence}}, }\href
  {\doibase 10.1103/PhysRevResearch.2.043197} {\bibfield  {journal} {\bibinfo
  {journal} {Phys. Rev. Res.}\ }\textbf {\bibinfo {volume} {2}},\ \bibinfo
  {pages} {043197} (\bibinfo {year} {2020})}\BibitemShut {NoStop}%
\bibitem [{\citenamefont {Lin}\ and\ \citenamefont
  {Hughes}(2018)}]{LinMao18PRB}%
  \BibitemOpen
  \bibfield  {author} {\bibinfo {author} {\bibfnamefont {M.}~\bibnamefont
  {Lin}}\ and\ \bibinfo {author} {\bibfnamefont {T.~L.}\ \bibnamefont
  {Hughes}},\ }\bibfield  {title} {\enquote {\bibinfo {title} {Topological
  quadrupolar semimetals}}, }\href {\doibase 10.1103/PhysRevB.98.241103}
  {\bibfield  {journal} {\bibinfo  {journal} {Phys. Rev. B}\ }\textbf {\bibinfo
  {volume} {98}},\ \bibinfo {pages} {241103} (\bibinfo {year}
  {2018})}\BibitemShut {NoStop}%
\bibitem [{\citenamefont {Chen}\ \emph
  {et~al.}(2021{\natexlab{a}})\citenamefont {Chen}, \citenamefont {Liu},
  \citenamefont {Wang}, \citenamefont {Lu},\ and\ \citenamefont
  {Xie}}]{ChenR21PRL}%
  \BibitemOpen
  \bibfield  {author} {\bibinfo {author} {\bibfnamefont {R.}~\bibnamefont
  {Chen}}, \bibinfo {author} {\bibfnamefont {T.}~\bibnamefont {Liu}}, \bibinfo
  {author} {\bibfnamefont {C.~M.}\ \bibnamefont {Wang}}, \bibinfo {author}
  {\bibfnamefont {H.-Z.}\ \bibnamefont {Lu}}, \ and\ \bibinfo {author}
  {\bibfnamefont {X.~C.}\ \bibnamefont {Xie}},\ }\bibfield  {title} {\enquote
  {\bibinfo {title} {Field-tunable one-sided higher-order topological hinge
  states in {Dirac} semimetals}}, }\href {\doibase
  10.1103/PhysRevLett.127.066801} {\bibfield  {journal} {\bibinfo  {journal}
  {Phys. Rev. Lett.}\ }\textbf {\bibinfo {volume} {127}},\ \bibinfo {pages}
  {066801} (\bibinfo {year} {2021}{\natexlab{a}})}\BibitemShut {NoStop}%
\bibitem [{\citenamefont {Wang}\ \emph
  {et~al.}(2022{\natexlab{a}})\citenamefont {Wang}, \citenamefont {Wang},
  \citenamefont {Sun}, \citenamefont {Chen},\ and\ \citenamefont
  {Xu}}]{WangZM22arXiv}%
  \BibitemOpen
  \bibfield  {author} {\bibinfo {author} {\bibfnamefont {Z.-M.}\ \bibnamefont
  {Wang}}, \bibinfo {author} {\bibfnamefont {R.}~\bibnamefont {Wang}}, \bibinfo
  {author} {\bibfnamefont {J.-H.}\ \bibnamefont {Sun}}, \bibinfo {author}
  {\bibfnamefont {T.-Y.}\ \bibnamefont {Chen}}, \ and\ \bibinfo {author}
  {\bibfnamefont {D.-H.}\ \bibnamefont {Xu}},\ }\bibfield  {title} {\enquote
  {\bibinfo {title} {{Floquet Weyl semimetal phases in light-irradiated
  higher-order topological Dirac semimetals}}}, }\href
  {https://arxiv.org/abs/2210.01012} {\bibfield  {journal} {\bibinfo  {journal}
  {arXiv: 2210.01012}\ } (\bibinfo {year} {2022}{\natexlab{a}})}\BibitemShut
  {NoStop}%
\bibitem [{\citenamefont {Wang}\ \emph {et~al.}(2020)\citenamefont {Wang},
  \citenamefont {Dai}, \citenamefont {Shao}, \citenamefont {Yang},\ and\
  \citenamefont {Zhao}}]{Wang20PRL}%
  \BibitemOpen
  \bibfield  {author} {\bibinfo {author} {\bibfnamefont {K.}~\bibnamefont
  {Wang}}, \bibinfo {author} {\bibfnamefont {J.-X.}\ \bibnamefont {Dai}},
  \bibinfo {author} {\bibfnamefont {L.~B.}\ \bibnamefont {Shao}}, \bibinfo
  {author} {\bibfnamefont {S.~A.}\ \bibnamefont {Yang}}, \ and\ \bibinfo
  {author} {\bibfnamefont {Y.~X.}\ \bibnamefont {Zhao}},\ }\bibfield  {title}
  {\enquote {\bibinfo {title} {Boundary criticality of $\mathcal{PT}$-invariant
  topology and second-order nodal-line semimetals}}, }\href {\doibase
  10.1103/PhysRevLett.125.126403} {\bibfield  {journal} {\bibinfo  {journal}
  {Phys. Rev. Lett.}\ }\textbf {\bibinfo {volume} {125}},\ \bibinfo {pages}
  {126403} (\bibinfo {year} {2020})}\BibitemShut {NoStop}%
\bibitem [{\citenamefont {Tran}\ \emph {et~al.}(2015)\citenamefont {Tran},
  \citenamefont {Dauphin}, \citenamefont {Goldman},\ and\ \citenamefont
  {Gaspard}}]{Tran2015PRB}%
  \BibitemOpen
  \bibfield  {author} {\bibinfo {author} {\bibfnamefont {D.-T.}\ \bibnamefont
  {Tran}}, \bibinfo {author} {\bibfnamefont {A.}~\bibnamefont {Dauphin}},
  \bibinfo {author} {\bibfnamefont {N.}~\bibnamefont {Goldman}}, \ and\
  \bibinfo {author} {\bibfnamefont {P.}~\bibnamefont {Gaspard}},\ }\bibfield
  {title} {\enquote {\bibinfo {title} {{Topological Hofstadter insulators in a
  two-dimensional quasicrystal}}}, }\href {\doibase 10.1103/physrevb.91.085125}
  {\bibfield  {journal} {\bibinfo  {journal} {Phys. Rev. B}\ }\textbf {\bibinfo
  {volume} {91}},\ \bibinfo {pages} {085125} (\bibinfo {year}
  {2015})}\BibitemShut {NoStop}%
\bibitem [{\citenamefont {Duncan}\ \emph {et~al.}(2020)\citenamefont {Duncan},
  \citenamefont {Manna},\ and\ \citenamefont {Nielsen}}]{Duncan20PRB}%
  \BibitemOpen
  \bibfield  {author} {\bibinfo {author} {\bibfnamefont {C.~W.}\ \bibnamefont
  {Duncan}}, \bibinfo {author} {\bibfnamefont {S.}~\bibnamefont {Manna}}, \
  and\ \bibinfo {author} {\bibfnamefont {A.~E.~B.}\ \bibnamefont {Nielsen}},\
  }\bibfield  {title} {\enquote {\bibinfo {title} {Topological models in
  rotationally symmetric quasicrystals}}, }\href {\doibase
  10.1103/PhysRevB.101.115413} {\bibfield  {journal} {\bibinfo  {journal}
  {Phys. Rev. B}\ }\textbf {\bibinfo {volume} {101}},\ \bibinfo {pages}
  {115413} (\bibinfo {year} {2020})}\BibitemShut {NoStop}%
\bibitem [{\citenamefont {Li}\ and\ \citenamefont {Wang}(2020)}]{Li2020CPB}%
  \BibitemOpen
  \bibfield  {author} {\bibinfo {author} {\bibfnamefont {Z.}~\bibnamefont
  {Li}}\ and\ \bibinfo {author} {\bibfnamefont {Z.~F.}\ \bibnamefont {Wang}},\
  }\bibfield  {title} {\enquote {\bibinfo {title} {Quantum anomalous hall
  effect in twisted bilayer graphene quasicrystal}}, }\href {\doibase
  10.1088/1674-1056/abab77} {\bibfield  {journal} {\bibinfo  {journal} {Chin.
  Phys. B}\ }\textbf {\bibinfo {volume} {29}},\ \bibinfo {pages} {107101}
  (\bibinfo {year} {2020})}\BibitemShut {NoStop}%
\bibitem [{\citenamefont {Jeon}\ \emph {et~al.}(2022)\citenamefont {Jeon},
  \citenamefont {Park},\ and\ \citenamefont {Lee}}]{Jeon22PRB}%
  \BibitemOpen
  \bibfield  {author} {\bibinfo {author} {\bibfnamefont {J.}~\bibnamefont
  {Jeon}}, \bibinfo {author} {\bibfnamefont {M.~J.}\ \bibnamefont {Park}}, \
  and\ \bibinfo {author} {\bibfnamefont {S.}~\bibnamefont {Lee}},\ }\bibfield
  {title} {\enquote {\bibinfo {title} {{Length scale formation in the Landau
  levels of quasicrystals}}}, }\href {\doibase 10.1103/PhysRevB.105.045146}
  {\bibfield  {journal} {\bibinfo  {journal} {Phys. Rev. B}\ }\textbf {\bibinfo
  {volume} {105}},\ \bibinfo {pages} {045146} (\bibinfo {year}
  {2022})}\BibitemShut {NoStop}%
\bibitem [{\citenamefont {He}\ \emph {et~al.}(2019)\citenamefont {He},
  \citenamefont {Ding}, \citenamefont {Zhou}, \citenamefont {Wang},\ and\
  \citenamefont {Gong}}]{He19PRB}%
  \BibitemOpen
  \bibfield  {author} {\bibinfo {author} {\bibfnamefont {A.-L.}\ \bibnamefont
  {He}}, \bibinfo {author} {\bibfnamefont {L.-R.}\ \bibnamefont {Ding}},
  \bibinfo {author} {\bibfnamefont {Y.}~\bibnamefont {Zhou}}, \bibinfo {author}
  {\bibfnamefont {Y.-F.}\ \bibnamefont {Wang}}, \ and\ \bibinfo {author}
  {\bibfnamefont {C.-D.}\ \bibnamefont {Gong}},\ }\bibfield  {title} {\enquote
  {\bibinfo {title} {{Quasicrystalline Chern insulators}}}, }\href {\doibase
  10.1103/PhysRevB.100.214109} {\bibfield  {journal} {\bibinfo  {journal}
  {Phys. Rev. B}\ }\textbf {\bibinfo {volume} {100}},\ \bibinfo {pages}
  {214109} (\bibinfo {year} {2019})}\BibitemShut {NoStop}%
\bibitem [{\citenamefont {Traverso}\ \emph {et~al.}(2022)\citenamefont
  {Traverso}, \citenamefont {Sassetti},\ and\ \citenamefont
  {Ziani}}]{Traverso22PRB}%
  \BibitemOpen
  \bibfield  {author} {\bibinfo {author} {\bibfnamefont {S.}~\bibnamefont
  {Traverso}}, \bibinfo {author} {\bibfnamefont {M.}~\bibnamefont {Sassetti}},
  \ and\ \bibinfo {author} {\bibfnamefont {N.~T.}\ \bibnamefont {Ziani}},\
  }\bibfield  {title} {\enquote {\bibinfo {title} {{Role of the edges in a
  quasicrystalline Haldane model}}}, }\href {\doibase
  10.1103/PhysRevB.106.125428} {\bibfield  {journal} {\bibinfo  {journal}
  {Phys. Rev. B}\ }\textbf {\bibinfo {volume} {106}},\ \bibinfo {pages}
  {125428} (\bibinfo {year} {2022})}\BibitemShut {NoStop}%
\bibitem [{\citenamefont {Huang}\ and\ \citenamefont {Liu}(2018)}]{Huang18PRL}%
  \BibitemOpen
  \bibfield  {author} {\bibinfo {author} {\bibfnamefont {H.}~\bibnamefont
  {Huang}}\ and\ \bibinfo {author} {\bibfnamefont {F.}~\bibnamefont {Liu}},\
  }\bibfield  {title} {\enquote {\bibinfo {title} {Quantum spin {Hall} effect
  and spin {Bott} index in a quasicrystal lattice}}, }\href {\doibase
  10.1103/PhysRevLett.121.126401} {\bibfield  {journal} {\bibinfo  {journal}
  {Phys. Rev. Lett.}\ }\textbf {\bibinfo {volume} {121}},\ \bibinfo {pages}
  {126401} (\bibinfo {year} {2018})}\BibitemShut {NoStop}%
\bibitem [{\citenamefont {Spurrier}\ and\ \citenamefont
  {Cooper}(2020{\natexlab{a}})}]{Spurrier20PRR}%
  \BibitemOpen
  \bibfield  {author} {\bibinfo {author} {\bibfnamefont {S.}~\bibnamefont
  {Spurrier}}\ and\ \bibinfo {author} {\bibfnamefont {N.~R.}\ \bibnamefont
  {Cooper}},\ }\bibfield  {title} {\enquote {\bibinfo {title} {Kane-mele with a
  twist: Quasicrystalline higher-order topological insulators with fractional
  mass kinks}}, }\href {\doibase 10.1103/PhysRevResearch.2.033071} {\bibfield
  {journal} {\bibinfo  {journal} {Phys. Rev. Res.}\ }\textbf {\bibinfo {volume}
  {2}},\ \bibinfo {pages} {033071} (\bibinfo {year}
  {2020}{\natexlab{a}})}\BibitemShut {NoStop}%
\bibitem [{\citenamefont {Bandres}\ \emph {et~al.}(2016)\citenamefont
  {Bandres}, \citenamefont {Rechtsman},\ and\ \citenamefont
  {Segev}}]{Bandres2016PRX}%
  \BibitemOpen
  \bibfield  {author} {\bibinfo {author} {\bibfnamefont {M.~A.}\ \bibnamefont
  {Bandres}}, \bibinfo {author} {\bibfnamefont {M.~C.}\ \bibnamefont
  {Rechtsman}}, \ and\ \bibinfo {author} {\bibfnamefont {M.}~\bibnamefont
  {Segev}},\ }\bibfield  {title} {\enquote {\bibinfo {title} {Topological
  photonic quasicrystals: Fractal topological spectrum and protected
  transport}}, }\href {\doibase 10.1103/physrevx.6.011016} {\bibfield
  {journal} {\bibinfo  {journal} {Phys. Rev. X}\ }\textbf {\bibinfo {volume}
  {6}},\ \bibinfo {pages} {011016} (\bibinfo {year} {2016})}\BibitemShut
  {NoStop}%
\bibitem [{\citenamefont {P\"oyh\"onen}\ \emph {et~al.}(2018)\citenamefont
  {P\"oyh\"onen}, \citenamefont {Sahlberg}, \citenamefont {Weststr\"om},\ and\
  \citenamefont {Ojanen}}]{Poyhonen2018NatCom}%
  \BibitemOpen
  \bibfield  {author} {\bibinfo {author} {\bibfnamefont {K.}~\bibnamefont
  {P\"oyh\"onen}}, \bibinfo {author} {\bibfnamefont {I.}~\bibnamefont
  {Sahlberg}}, \bibinfo {author} {\bibfnamefont {A.}~\bibnamefont
  {Weststr\"om}}, \ and\ \bibinfo {author} {\bibfnamefont {T.}~\bibnamefont
  {Ojanen}},\ }\bibfield  {title} {\enquote {\bibinfo {title} {Amorphous
  topological superconductivity in a shiba glass}}, }\href {\doibase
  10.1038/s41467-018-04532-x} {\bibfield  {journal} {\bibinfo  {journal} {Nat.
  Commun.}\ }\textbf {\bibinfo {volume} {9}},\ \bibinfo {pages} {2103}
  (\bibinfo {year} {2018})}\BibitemShut {NoStop}%
\bibitem [{\citenamefont {Fulga}\ \emph {et~al.}(2016)\citenamefont {Fulga},
  \citenamefont {Pikulin},\ and\ \citenamefont {Loring}}]{Fulga2016PRL}%
  \BibitemOpen
  \bibfield  {author} {\bibinfo {author} {\bibfnamefont {I.}~\bibnamefont
  {Fulga}}, \bibinfo {author} {\bibfnamefont {D.}~\bibnamefont {Pikulin}}, \
  and\ \bibinfo {author} {\bibfnamefont {T.}~\bibnamefont {Loring}},\
  }\bibfield  {title} {\enquote {\bibinfo {title} {Aperiodic weak topological
  superconductors}}, }\href {\doibase 10.1103/physrevlett.116.257002}
  {\bibfield  {journal} {\bibinfo  {journal} {Phys. Rev. Lett.}\ }\textbf
  {\bibinfo {volume} {116}},\ \bibinfo {pages} {257002} (\bibinfo {year}
  {2016})}\BibitemShut {NoStop}%
\bibitem [{\citenamefont {Manna}\ \emph {et~al.}(2022)\citenamefont {Manna},
  \citenamefont {Das},\ and\ \citenamefont {Roy}}]{Manna2022arXiv}%
  \BibitemOpen
  \bibfield  {author} {\bibinfo {author} {\bibfnamefont {S.}~\bibnamefont
  {Manna}}, \bibinfo {author} {\bibfnamefont {S.~K.}\ \bibnamefont {Das}}, \
  and\ \bibinfo {author} {\bibfnamefont {B.}~\bibnamefont {Roy}},\ }\bibfield
  {title} {\enquote {\bibinfo {title} {Noncrystalline topological
  superconductors}}, }\href {https://arxiv.org/abs/2207.02203} {\bibfield
  {journal} {\bibinfo  {journal} {arXiv:2207.02203}\ } (\bibinfo {year}
  {2022})}\BibitemShut {NoStop}%
\bibitem [{\citenamefont {Longhi}(2019)}]{Longhi19PRL}%
  \BibitemOpen
  \bibfield  {author} {\bibinfo {author} {\bibfnamefont {S.}~\bibnamefont
  {Longhi}},\ }\bibfield  {title} {\enquote {\bibinfo {title} {Topological
  phase transition in non-hermitian quasicrystals}}, }\href {\doibase
  10.1103/PhysRevLett.122.237601} {\bibfield  {journal} {\bibinfo  {journal}
  {Phys. Rev. Lett.}\ }\textbf {\bibinfo {volume} {122}},\ \bibinfo {pages}
  {237601} (\bibinfo {year} {2019})}\BibitemShut {NoStop}%
\bibitem [{\citenamefont {Zeng}\ \emph {et~al.}(2020)\citenamefont {Zeng},
  \citenamefont {Yang},\ and\ \citenamefont {Xu}}]{Zeng20PRB}%
  \BibitemOpen
  \bibfield  {author} {\bibinfo {author} {\bibfnamefont {Q.-B.}\ \bibnamefont
  {Zeng}}, \bibinfo {author} {\bibfnamefont {Y.-B.}\ \bibnamefont {Yang}}, \
  and\ \bibinfo {author} {\bibfnamefont {Y.}~\bibnamefont {Xu}},\ }\bibfield
  {title} {\enquote {\bibinfo {title} {{Topological phases in non-Hermitian
  Aubry-Andr\'e-Harper models}}}, }\href {\doibase 10.1103/PhysRevB.101.020201}
  {\bibfield  {journal} {\bibinfo  {journal} {Phys. Rev. B}\ }\textbf {\bibinfo
  {volume} {101}},\ \bibinfo {pages} {020201} (\bibinfo {year}
  {2020})}\BibitemShut {NoStop}%
\bibitem [{\citenamefont {Chen}\ \emph {et~al.}(2020)\citenamefont {Chen},
  \citenamefont {Chen}, \citenamefont {Gao}, \citenamefont {Zhou},\ and\
  \citenamefont {Xu}}]{ChenR20PRL}%
  \BibitemOpen
  \bibfield  {author} {\bibinfo {author} {\bibfnamefont {R.}~\bibnamefont
  {Chen}}, \bibinfo {author} {\bibfnamefont {C.-Z.}\ \bibnamefont {Chen}},
  \bibinfo {author} {\bibfnamefont {J.-H.}\ \bibnamefont {Gao}}, \bibinfo
  {author} {\bibfnamefont {B.}~\bibnamefont {Zhou}}, \ and\ \bibinfo {author}
  {\bibfnamefont {D.-H.}\ \bibnamefont {Xu}},\ }\bibfield  {title} {\enquote
  {\bibinfo {title} {Higher-order topological insulators in quasicrystals}},
  }\href {\doibase 10.1103/PhysRevLett.124.036803} {\bibfield  {journal}
  {\bibinfo  {journal} {Phys. Rev. Lett.}\ }\textbf {\bibinfo {volume} {124}},\
  \bibinfo {pages} {036803} (\bibinfo {year} {2020})}\BibitemShut {NoStop}%
\bibitem [{\citenamefont {Varjas}\ \emph {et~al.}(2019)\citenamefont {Varjas},
  \citenamefont {Lau}, \citenamefont {P\"oyh\"onen}, \citenamefont {Akhmerov},
  \citenamefont {Pikulin},\ and\ \citenamefont {Fulga}}]{Varjas19PRL}%
  \BibitemOpen
  \bibfield  {author} {\bibinfo {author} {\bibfnamefont {D.}~\bibnamefont
  {Varjas}}, \bibinfo {author} {\bibfnamefont {A.}~\bibnamefont {Lau}},
  \bibinfo {author} {\bibfnamefont {K.}~\bibnamefont {P\"oyh\"onen}}, \bibinfo
  {author} {\bibfnamefont {A.~R.}\ \bibnamefont {Akhmerov}}, \bibinfo {author}
  {\bibfnamefont {D.~I.}\ \bibnamefont {Pikulin}}, \ and\ \bibinfo {author}
  {\bibfnamefont {I.~C.}\ \bibnamefont {Fulga}},\ }\bibfield  {title} {\enquote
  {\bibinfo {title} {Topological phases without crystalline counterparts}},
  }\href {\doibase 10.1103/PhysRevLett.123.196401} {\bibfield  {journal}
  {\bibinfo  {journal} {Phys. Rev. Lett.}\ }\textbf {\bibinfo {volume} {123}},\
  \bibinfo {pages} {196401} (\bibinfo {year} {2019})}\BibitemShut {NoStop}%
\bibitem [{\citenamefont {Hua}\ \emph {et~al.}(2020)\citenamefont {Hua},
  \citenamefont {Chen}, \citenamefont {Zhou},\ and\ \citenamefont
  {Xu}}]{Hua20PRB}%
  \BibitemOpen
  \bibfield  {author} {\bibinfo {author} {\bibfnamefont {C.-B.}\ \bibnamefont
  {Hua}}, \bibinfo {author} {\bibfnamefont {R.}~\bibnamefont {Chen}}, \bibinfo
  {author} {\bibfnamefont {B.}~\bibnamefont {Zhou}}, \ and\ \bibinfo {author}
  {\bibfnamefont {D.-H.}\ \bibnamefont {Xu}},\ }\bibfield  {title} {\enquote
  {\bibinfo {title} {Higher-order topological insulator in a dodecagonal
  quasicrystal}}, }\href {\doibase 10.1103/PhysRevB.102.241102} {\bibfield
  {journal} {\bibinfo  {journal} {Phys. Rev. B}\ }\textbf {\bibinfo {volume}
  {102}},\ \bibinfo {pages} {241102} (\bibinfo {year} {2020})}\BibitemShut
  {NoStop}%
\bibitem [{\citenamefont {Spurrier}\ and\ \citenamefont
  {Cooper}(2020{\natexlab{b}})}]{PhysRevResearch.2.033071}%
  \BibitemOpen
  \bibfield  {author} {\bibinfo {author} {\bibfnamefont {S.}~\bibnamefont
  {Spurrier}}\ and\ \bibinfo {author} {\bibfnamefont {N.~R.}\ \bibnamefont
  {Cooper}},\ }\bibfield  {title} {\enquote {\bibinfo {title} {Kane-mele with a
  twist: Quasicrystalline higher-order topological insulators with fractional
  mass kinks}}, }\href {\doibase 10.1103/PhysRevResearch.2.033071} {\bibfield
  {journal} {\bibinfo  {journal} {Phys. Rev. Res.}\ }\textbf {\bibinfo {volume}
  {2}},\ \bibinfo {pages} {033071} (\bibinfo {year}
  {2020}{\natexlab{b}})}\BibitemShut {NoStop}%
\bibitem [{\citenamefont {Huang}\ \emph {et~al.}(2021)\citenamefont {Huang},
  \citenamefont {Fan}, \citenamefont {Li},\ and\ \citenamefont
  {Liu}}]{Huang2021NanoLetter}%
  \BibitemOpen
  \bibfield  {author} {\bibinfo {author} {\bibfnamefont {H.}~\bibnamefont
  {Huang}}, \bibinfo {author} {\bibfnamefont {J.}~\bibnamefont {Fan}}, \bibinfo
  {author} {\bibfnamefont {D.}~\bibnamefont {Li}}, \ and\ \bibinfo {author}
  {\bibfnamefont {F.}~\bibnamefont {Liu}},\ }\bibfield  {title} {\enquote
  {\bibinfo {title} {Generic orbital design of higher-order topological
  quasicrystalline insulators with odd five-fold rotation symmetry}}, }\href
  {\doibase 10.1021/acs.nanolett.1c02661} {\bibfield  {journal} {\bibinfo
  {journal} {Nano Lett.}\ }\textbf {\bibinfo {volume} {21}},\ \bibinfo {pages}
  {7056} (\bibinfo {year} {2021})}\BibitemShut {NoStop}%
\bibitem [{\citenamefont {Wang}\ \emph
  {et~al.}(2022{\natexlab{b}})\citenamefont {Wang}, \citenamefont {Liu},\ and\
  \citenamefont {Huang}}]{Wang22PRL}%
  \BibitemOpen
  \bibfield  {author} {\bibinfo {author} {\bibfnamefont {C.}~\bibnamefont
  {Wang}}, \bibinfo {author} {\bibfnamefont {F.}~\bibnamefont {Liu}}, \ and\
  \bibinfo {author} {\bibfnamefont {H.}~\bibnamefont {Huang}},\ }\bibfield
  {title} {\enquote {\bibinfo {title} {Effective model for fractional
  topological corner modes in quasicrystals}}, }\href {\doibase
  10.1103/PhysRevLett.129.056403} {\bibfield  {journal} {\bibinfo  {journal}
  {Phys. Rev. Lett.}\ }\textbf {\bibinfo {volume} {129}},\ \bibinfo {pages}
  {056403} (\bibinfo {year} {2022}{\natexlab{b}})}\BibitemShut {NoStop}%
\bibitem [{\citenamefont {Shi}\ \emph {et~al.}(2022)\citenamefont {Shi},
  \citenamefont {Jiang}, \citenamefont {Peng}, \citenamefont {Peng},
  \citenamefont {Chen},\ and\ \citenamefont {Liu}}]{Shi2022arXiv}%
  \BibitemOpen
  \bibfield  {author} {\bibinfo {author} {\bibfnamefont {A.}~\bibnamefont
  {Shi}}, \bibinfo {author} {\bibfnamefont {J.}~\bibnamefont {Jiang}}, \bibinfo
  {author} {\bibfnamefont {Y.}~\bibnamefont {Peng}}, \bibinfo {author}
  {\bibfnamefont {P.}~\bibnamefont {Peng}}, \bibinfo {author} {\bibfnamefont
  {J.}~\bibnamefont {Chen}}, \ and\ \bibinfo {author} {\bibfnamefont
  {J.}~\bibnamefont {Liu}},\ }\bibfield  {title} {\enquote {\bibinfo {title}
  {Multimer analysis method reveals higher-order topology in quasicrystals}},
  }\href {https://arxiv.org/ftp/arxiv/papers/2209/2209.05751.pdf} {\bibfield
  {journal} {\bibinfo  {journal} {arXiv:2209.05751}\ } (\bibinfo {year}
  {2022})}\BibitemShut {NoStop}%
\bibitem [{\citenamefont {Xiong}\ \emph {et~al.}(2022)\citenamefont {Xiong},
  \citenamefont {Zhang}, \citenamefont {Liu}, \citenamefont {Zheng},\ and\
  \citenamefont {Jiang}}]{Xiong2022arXiv}%
  \BibitemOpen
  \bibfield  {author} {\bibinfo {author} {\bibfnamefont {L.}~\bibnamefont
  {Xiong}}, \bibinfo {author} {\bibfnamefont {Y.}~\bibnamefont {Zhang}},
  \bibinfo {author} {\bibfnamefont {Y.}~\bibnamefont {Liu}}, \bibinfo {author}
  {\bibfnamefont {Y.}~\bibnamefont {Zheng}}, \ and\ \bibinfo {author}
  {\bibfnamefont {X.}~\bibnamefont {Jiang}},\ }\bibfield  {title} {\enquote
  {\bibinfo {title} {Higher-order topological states in photonic thue-morse
  quasicrystals: quadrupole insulator and a new origin of corner states}},
  }\href {https://arxiv.org/abs/2207.12286} {\bibfield  {journal} {\bibinfo
  {journal} {arXiv:2207.12286}\ } (\bibinfo {year} {2022})}\BibitemShut
  {NoStop}%
\bibitem [{\citenamefont {Ghadimi}\ \emph {et~al.}(2021)\citenamefont
  {Ghadimi}, \citenamefont {Sugimoto}, \citenamefont {Tanaka},\ and\
  \citenamefont {Tohyama}}]{Ghadimi21PRB}%
  \BibitemOpen
  \bibfield  {author} {\bibinfo {author} {\bibfnamefont {R.}~\bibnamefont
  {Ghadimi}}, \bibinfo {author} {\bibfnamefont {T.}~\bibnamefont {Sugimoto}},
  \bibinfo {author} {\bibfnamefont {K.}~\bibnamefont {Tanaka}}, \ and\ \bibinfo
  {author} {\bibfnamefont {T.}~\bibnamefont {Tohyama}},\ }\bibfield  {title}
  {\enquote {\bibinfo {title} {Topological superconductivity in
  quasicrystals}}, }\href {\doibase 10.1103/PhysRevB.104.144511} {\bibfield
  {journal} {\bibinfo  {journal} {Phys. Rev. B}\ }\textbf {\bibinfo {volume}
  {104}},\ \bibinfo {pages} {144511} (\bibinfo {year} {2021})}\BibitemShut
  {NoStop}%
\bibitem [{\citenamefont {Lv}\ \emph {et~al.}(2021)\citenamefont {Lv},
  \citenamefont {Chen}, \citenamefont {Li}, \citenamefont {Guan}, \citenamefont
  {Zhou}, \citenamefont {Dong}, \citenamefont {Zhao}, \citenamefont {Li},
  \citenamefont {Wang}, \citenamefont {Tao}, \citenamefont {Shi},\ and\
  \citenamefont {Xu}}]{Lv2021ComPhys}%
  \BibitemOpen
  \bibfield  {author} {\bibinfo {author} {\bibfnamefont {B.}~\bibnamefont
  {Lv}}, \bibinfo {author} {\bibfnamefont {R.}~\bibnamefont {Chen}}, \bibinfo
  {author} {\bibfnamefont {R.}~\bibnamefont {Li}}, \bibinfo {author}
  {\bibfnamefont {C.}~\bibnamefont {Guan}}, \bibinfo {author} {\bibfnamefont
  {B.}~\bibnamefont {Zhou}}, \bibinfo {author} {\bibfnamefont {G.}~\bibnamefont
  {Dong}},  \emph {et~al.},\ }\bibfield  {title} {\enquote {\bibinfo {title}
  {Realization of quasicrystalline quadrupole topological insulators in
  electrical circuits}}, }\href {\doibase 10.1038/s42005-021-00610-7}
  {\bibfield  {journal} {\bibinfo  {journal} {Commun. Phys.}\ }\textbf
  {\bibinfo {volume} {4}},\ \bibinfo {pages} {108} (\bibinfo {year}
  {2021})}\BibitemShut {NoStop}%
\bibitem [{\citenamefont {Chen}\ \emph {et~al.}(2019)\citenamefont {Chen},
  \citenamefont {Xu},\ and\ \citenamefont {Zhou}}]{ChenR19PRB}%
  \BibitemOpen
  \bibfield  {author} {\bibinfo {author} {\bibfnamefont {R.}~\bibnamefont
  {Chen}}, \bibinfo {author} {\bibfnamefont {D.-H.}\ \bibnamefont {Xu}}, \ and\
  \bibinfo {author} {\bibfnamefont {B.}~\bibnamefont {Zhou}},\ }\bibfield
  {title} {\enquote {\bibinfo {title} {Topological {Anderson} insulator phase
  in a quasicrystal lattice}}, }\href {\doibase 10.1103/PhysRevB.100.115311}
  {\bibfield  {journal} {\bibinfo  {journal} {Phys. Rev. B}\ }\textbf {\bibinfo
  {volume} {100}},\ \bibinfo {pages} {115311} (\bibinfo {year}
  {2019})}\BibitemShut {NoStop}%
\bibitem [{\citenamefont {Hua}\ \emph {et~al.}(2021)\citenamefont {Hua},
  \citenamefont {Liu}, \citenamefont {Peng}, \citenamefont {Chen},
  \citenamefont {Xu},\ and\ \citenamefont {Zhou}}]{Hua21PRB}%
  \BibitemOpen
  \bibfield  {author} {\bibinfo {author} {\bibfnamefont {C.-B.}\ \bibnamefont
  {Hua}}, \bibinfo {author} {\bibfnamefont {Z.-R.}\ \bibnamefont {Liu}},
  \bibinfo {author} {\bibfnamefont {T.}~\bibnamefont {Peng}}, \bibinfo {author}
  {\bibfnamefont {R.}~\bibnamefont {Chen}}, \bibinfo {author} {\bibfnamefont
  {D.-H.}\ \bibnamefont {Xu}}, \ and\ \bibinfo {author} {\bibfnamefont
  {B.}~\bibnamefont {Zhou}},\ }\bibfield  {title} {\enquote {\bibinfo {title}
  {{Disorder-induced chiral and helical Majorana edge modes in a
  two-dimensional Ammann-Beenker quasicrystal}}}, }\href {\doibase
  10.1103/PhysRevB.104.155304} {\bibfield  {journal} {\bibinfo  {journal}
  {Phys. Rev. B}\ }\textbf {\bibinfo {volume} {104}},\ \bibinfo {pages}
  {155304} (\bibinfo {year} {2021})}\BibitemShut {NoStop}%
\bibitem [{\citenamefont {Peng}\ \emph
  {et~al.}(2021{\natexlab{a}})\citenamefont {Peng}, \citenamefont {Hua},
  \citenamefont {Chen}, \citenamefont {Xu},\ and\ \citenamefont
  {Zhou}}]{Peng21PRB}%
  \BibitemOpen
  \bibfield  {author} {\bibinfo {author} {\bibfnamefont {T.}~\bibnamefont
  {Peng}}, \bibinfo {author} {\bibfnamefont {C.-B.}\ \bibnamefont {Hua}},
  \bibinfo {author} {\bibfnamefont {R.}~\bibnamefont {Chen}}, \bibinfo {author}
  {\bibfnamefont {D.-H.}\ \bibnamefont {Xu}}, \ and\ \bibinfo {author}
  {\bibfnamefont {B.}~\bibnamefont {Zhou}},\ }\bibfield  {title} {\enquote
  {\bibinfo {title} {{Topological Anderson insulators in an Ammann-Beenker
  quasicrystal and a snub-square crystal}}}, }\href {\doibase
  10.1103/PhysRevB.103.085307} {\bibfield  {journal} {\bibinfo  {journal}
  {Phys. Rev. B}\ }\textbf {\bibinfo {volume} {103}},\ \bibinfo {pages}
  {085307} (\bibinfo {year} {2021}{\natexlab{a}})}\BibitemShut {NoStop}%
\bibitem [{\citenamefont {Peng}\ \emph
  {et~al.}(2021{\natexlab{b}})\citenamefont {Peng}, \citenamefont {Hua},
  \citenamefont {Chen}, \citenamefont {Liu}, \citenamefont {Xu},\ and\
  \citenamefont {Zhou}}]{Peng21PRB1}%
  \BibitemOpen
  \bibfield  {author} {\bibinfo {author} {\bibfnamefont {T.}~\bibnamefont
  {Peng}}, \bibinfo {author} {\bibfnamefont {C.-B.}\ \bibnamefont {Hua}},
  \bibinfo {author} {\bibfnamefont {R.}~\bibnamefont {Chen}}, \bibinfo {author}
  {\bibfnamefont {Z.-R.}\ \bibnamefont {Liu}}, \bibinfo {author} {\bibfnamefont
  {D.-H.}\ \bibnamefont {Xu}}, \ and\ \bibinfo {author} {\bibfnamefont
  {B.}~\bibnamefont {Zhou}},\ }\bibfield  {title} {\enquote {\bibinfo {title}
  {{Higher-order topological Anderson insulators in quasicrystals}}}, }\href
  {\doibase 10.1103/PhysRevB.104.245302} {\bibfield  {journal} {\bibinfo
  {journal} {Phys. Rev. B}\ }\textbf {\bibinfo {volume} {104}},\ \bibinfo
  {pages} {245302} (\bibinfo {year} {2021}{\natexlab{b}})}\BibitemShut
  {NoStop}%
\bibitem [{\citenamefont {Verbin}\ \emph {et~al.}(2013)\citenamefont {Verbin},
  \citenamefont {Zilberberg}, \citenamefont {Kraus}, \citenamefont {Lahini},\
  and\ \citenamefont {Silberberg}}]{Verbin2013PRL}%
  \BibitemOpen
  \bibfield  {author} {\bibinfo {author} {\bibfnamefont {M.}~\bibnamefont
  {Verbin}}, \bibinfo {author} {\bibfnamefont {O.}~\bibnamefont {Zilberberg}},
  \bibinfo {author} {\bibfnamefont {Y.~E.}\ \bibnamefont {Kraus}}, \bibinfo
  {author} {\bibfnamefont {Y.}~\bibnamefont {Lahini}}, \ and\ \bibinfo {author}
  {\bibfnamefont {Y.}~\bibnamefont {Silberberg}},\ }\bibfield  {title}
  {\enquote {\bibinfo {title} {Observation of topological phase transitions in
  photonic quasicrystals}}, }\href {\doibase 10.1103/physrevlett.110.076403}
  {\bibfield  {journal} {\bibinfo  {journal} {Phys. Rev. Lett.}\ }\textbf
  {\bibinfo {volume} {110}},\ \bibinfo {pages} {076403} (\bibinfo {year}
  {2013})}\BibitemShut {NoStop}%
\bibitem [{\citenamefont {Apigo}\ \emph {et~al.}(2019)\citenamefont {Apigo},
  \citenamefont {Cheng}, \citenamefont {Dobiszewski}, \citenamefont {Prodan},\
  and\ \citenamefont {Prodan}}]{Apigo2019PRL}%
  \BibitemOpen
  \bibfield  {author} {\bibinfo {author} {\bibfnamefont {D.~J.}\ \bibnamefont
  {Apigo}}, \bibinfo {author} {\bibfnamefont {W.}~\bibnamefont {Cheng}},
  \bibinfo {author} {\bibfnamefont {K.~F.}\ \bibnamefont {Dobiszewski}},
  \bibinfo {author} {\bibfnamefont {E.}~\bibnamefont {Prodan}}, \ and\ \bibinfo
  {author} {\bibfnamefont {C.}~\bibnamefont {Prodan}},\ }\bibfield  {title}
  {\enquote {\bibinfo {title} {Observation of topological edge modes in a
  quasiperiodic acoustic waveguide}}, }\href {\doibase
  10.1103/physrevlett.122.095501} {\bibfield  {journal} {\bibinfo  {journal}
  {Phys. Rev. Lett.}\ }\textbf {\bibinfo {volume} {122}},\ \bibinfo {pages}
  {095501} (\bibinfo {year} {2019})}\BibitemShut {NoStop}%
\bibitem [{\citenamefont {Shechtman}\ \emph {et~al.}(1984)\citenamefont
  {Shechtman}, \citenamefont {Blech}, \citenamefont {Gratias},\ and\
  \citenamefont {Cahn}}]{Shechtman1984PRL}%
  \BibitemOpen
  \bibfield  {author} {\bibinfo {author} {\bibfnamefont {D.}~\bibnamefont
  {Shechtman}}, \bibinfo {author} {\bibfnamefont {I.}~\bibnamefont {Blech}},
  \bibinfo {author} {\bibfnamefont {D.}~\bibnamefont {Gratias}}, \ and\
  \bibinfo {author} {\bibfnamefont {J.~W.}\ \bibnamefont {Cahn}},\ }\bibfield
  {title} {\enquote {\bibinfo {title} {Metallic phase with long-range
  orientational order and no translational symmetry}}, }\href {\doibase
  10.1103/physrevlett.53.1951} {\bibfield  {journal} {\bibinfo  {journal}
  {Phys. Rev. Lett.}\ }\textbf {\bibinfo {volume} {53}},\ \bibinfo {pages}
  {1951} (\bibinfo {year} {1984})}\BibitemShut {NoStop}%
\bibitem [{\citenamefont {Levine}\ and\ \citenamefont
  {Steinhardt}(1984)}]{Levine1984PRL}%
  \BibitemOpen
  \bibfield  {author} {\bibinfo {author} {\bibfnamefont {D.}~\bibnamefont
  {Levine}}\ and\ \bibinfo {author} {\bibfnamefont {P.~J.}\ \bibnamefont
  {Steinhardt}},\ }\bibfield  {title} {\enquote {\bibinfo {title}
  {Quasicrystals: A new class of ordered structures}}, }\href {\doibase
  10.1103/physrevlett.53.2477} {\bibfield  {journal} {\bibinfo  {journal}
  {Phys. Rev. Lett.}\ }\textbf {\bibinfo {volume} {53}},\ \bibinfo {pages}
  {2477} (\bibinfo {year} {1984})}\BibitemShut {NoStop}%
\bibitem [{\citenamefont {Biggs}\ \emph {et~al.}(1990)\citenamefont {Biggs},
  \citenamefont {Poon},\ and\ \citenamefont {Munirathnam}}]{Biggs1990PRL}%
  \BibitemOpen
  \bibfield  {author} {\bibinfo {author} {\bibfnamefont {B.~D.}\ \bibnamefont
  {Biggs}}, \bibinfo {author} {\bibfnamefont {S.~J.}\ \bibnamefont {Poon}}, \
  and\ \bibinfo {author} {\bibfnamefont {N.~R.}\ \bibnamefont {Munirathnam}},\
  }\bibfield  {title} {\enquote {\bibinfo {title} {Stable al-cu-ru icosahedral
  crystals: A new class of electronic alloys}}, }\href {\doibase
  10.1103/physrevlett.65.2700} {\bibfield  {journal} {\bibinfo  {journal}
  {Phys. Rev. Lett.}\ }\textbf {\bibinfo {volume} {65}},\ \bibinfo {pages}
  {2700} (\bibinfo {year} {1990})}\BibitemShut {NoStop}%
\bibitem [{\citenamefont {Pierce}\ \emph {et~al.}(1993)\citenamefont {Pierce},
  \citenamefont {Poon},\ and\ \citenamefont {Guo}}]{Pierce1993Science}%
  \BibitemOpen
  \bibfield  {author} {\bibinfo {author} {\bibfnamefont {F.~S.}\ \bibnamefont
  {Pierce}}, \bibinfo {author} {\bibfnamefont {S.~J.}\ \bibnamefont {Poon}}, \
  and\ \bibinfo {author} {\bibfnamefont {Q.}~\bibnamefont {Guo}},\ }\bibfield
  {title} {\enquote {\bibinfo {title} {Electron localization in metallic
  quasicrystals}}, }\href {\doibase 10.1126/science.261.5122.737} {\bibfield
  {journal} {\bibinfo  {journal} {Science}\ }\textbf {\bibinfo {volume}
  {261}},\ \bibinfo {pages} {737} (\bibinfo {year} {1993})}\BibitemShut
  {NoStop}%
\bibitem [{\citenamefont {Basov}\ \emph {et~al.}(1994)\citenamefont {Basov},
  \citenamefont {Pierce}, \citenamefont {Volkov}, \citenamefont {Poon},\ and\
  \citenamefont {Timusk}}]{Basov1994PRL}%
  \BibitemOpen
  \bibfield  {author} {\bibinfo {author} {\bibfnamefont {D.~N.}\ \bibnamefont
  {Basov}}, \bibinfo {author} {\bibfnamefont {F.~S.}\ \bibnamefont {Pierce}},
  \bibinfo {author} {\bibfnamefont {P.}~\bibnamefont {Volkov}}, \bibinfo
  {author} {\bibfnamefont {S.~J.}\ \bibnamefont {Poon}}, \ and\ \bibinfo
  {author} {\bibfnamefont {T.}~\bibnamefont {Timusk}},\ }\bibfield  {title}
  {\enquote {\bibinfo {title} {Optical conductivity of insulating al-based
  alloys: Comparison of quasiperiodic and periodic systems}}, }\href {\doibase
  10.1103/physrevlett.73.1865} {\bibfield  {journal} {\bibinfo  {journal}
  {Phys. Rev. Lett.}\ }\textbf {\bibinfo {volume} {73}},\ \bibinfo {pages}
  {1865} (\bibinfo {year} {1994})}\BibitemShut {NoStop}%
\bibitem [{\citenamefont {Pierce}\ \emph {et~al.}(1994)\citenamefont {Pierce},
  \citenamefont {Guo},\ and\ \citenamefont {Poon}}]{Pierce1994PRL}%
  \BibitemOpen
  \bibfield  {author} {\bibinfo {author} {\bibfnamefont {F.~S.}\ \bibnamefont
  {Pierce}}, \bibinfo {author} {\bibfnamefont {Q.}~\bibnamefont {Guo}}, \ and\
  \bibinfo {author} {\bibfnamefont {S.~J.}\ \bibnamefont {Poon}},\ }\bibfield
  {title} {\enquote {\bibinfo {title} {Enhanced insulatorlike electron
  transport behavior of thermally tuned quasicrystalline states of al-pd-re
  alloys}}, }\href {\doibase 10.1103/physrevlett.73.2220} {\bibfield  {journal}
  {\bibinfo  {journal} {Phys. Rev. Lett.}\ }\textbf {\bibinfo {volume} {73}},\
  \bibinfo {pages} {2220} (\bibinfo {year} {1994})}\BibitemShut {NoStop}%
\bibitem [{\citenamefont {Cain}\ \emph {et~al.}(2020)\citenamefont {Cain},
  \citenamefont {Azizi}, \citenamefont {Conrad}, \citenamefont {Griffin},\ and\
  \citenamefont {Zettl}}]{Cain2020PNAS}%
  \BibitemOpen
  \bibfield  {author} {\bibinfo {author} {\bibfnamefont {J.~D.}\ \bibnamefont
  {Cain}}, \bibinfo {author} {\bibfnamefont {A.}~\bibnamefont {Azizi}},
  \bibinfo {author} {\bibfnamefont {M.}~\bibnamefont {Conrad}}, \bibinfo
  {author} {\bibfnamefont {S.~M.}\ \bibnamefont {Griffin}}, \ and\ \bibinfo
  {author} {\bibfnamefont {A.}~\bibnamefont {Zettl}},\ }\bibfield  {title}
  {\enquote {\bibinfo {title} {Layer-dependent topological phase in a
  two-dimensional quasicrystal and approximant}}, }\href {\doibase
  10.1073/pnas.2015164117} {\bibfield  {journal} {\bibinfo  {journal} {Proc.
  Natl. Acad. Sci. USA}\ }\textbf {\bibinfo {volume} {117}},\ \bibinfo {pages}
  {26135} (\bibinfo {year} {2020})}\BibitemShut {NoStop}%
\bibitem [{\citenamefont {Fonseca}\ \emph {et~al.}(2022)\citenamefont
  {Fonseca}, \citenamefont {Christensen}, \citenamefont {Joannopoulos},\ and\
  \citenamefont {Solja\v{c}i\'c}}]{Fonseca2022arXiv}%
  \BibitemOpen
  \bibfield  {author} {\bibinfo {author} {\bibfnamefont {A.~G.~e.}\
  \bibnamefont {Fonseca}}, \bibinfo {author} {\bibfnamefont {T.}~\bibnamefont
  {Christensen}}, \bibinfo {author} {\bibfnamefont {J.~D.}\ \bibnamefont
  {Joannopoulos}}, \ and\ \bibinfo {author} {\bibfnamefont {M.}~\bibnamefont
  {Solja\v{c}i\'c}},\ }\bibfield  {title} {\enquote {\bibinfo {title}
  {{Quasicrystalline Weyl points and dense Fermi-Bragg arcs}}}, }\href
  {https://arxiv.org/abs/2211.14299} {\bibfield  {journal} {\bibinfo  {journal}
  {arXiv:2211.14299}\ } (\bibinfo {year} {2022})}\BibitemShut {NoStop}%
\bibitem [{\citenamefont {Burkov}\ and\ \citenamefont
  {Balents}(2011)}]{Burkov2011PRL}%
  \BibitemOpen
  \bibfield  {author} {\bibinfo {author} {\bibfnamefont {A.~A.}\ \bibnamefont
  {Burkov}}\ and\ \bibinfo {author} {\bibfnamefont {L.}~\bibnamefont
  {Balents}},\ }\bibfield  {title} {\enquote {\bibinfo {title} {Weyl semimetal
  in a topological insulator multilayer}}, }\href {\doibase
  10.1103/physrevlett.107.127205} {\bibfield  {journal} {\bibinfo  {journal}
  {Phys. Rev. Lett.}\ }\textbf {\bibinfo {volume} {107}},\ \bibinfo {pages}
  {127205} (\bibinfo {year} {2011})}\BibitemShut {NoStop}%
\bibitem [{\citenamefont {Mogi}\ \emph
  {et~al.}(2017{\natexlab{a}})\citenamefont {Mogi}, \citenamefont {Kawamura},
  \citenamefont {Yoshimi}, \citenamefont {Tsukazaki}, \citenamefont {Kozuka},
  \citenamefont {Shirakawa}, \citenamefont {Takahashi}, \citenamefont
  {Kawasaki},\ and\ \citenamefont {Tokura}}]{Mogi17nm}%
  \BibitemOpen
  \bibfield  {author} {\bibinfo {author} {\bibfnamefont {M.}~\bibnamefont
  {Mogi}}, \bibinfo {author} {\bibfnamefont {M.}~\bibnamefont {Kawamura}},
  \bibinfo {author} {\bibfnamefont {R.}~\bibnamefont {Yoshimi}}, \bibinfo
  {author} {\bibfnamefont {A.}~\bibnamefont {Tsukazaki}}, \bibinfo {author}
  {\bibfnamefont {Y.}~\bibnamefont {Kozuka}}, \bibinfo {author} {\bibfnamefont
  {N.}~\bibnamefont {Shirakawa}}, \bibinfo {author} {\bibfnamefont {K.~S.}\
  \bibnamefont {Takahashi}}, \bibinfo {author} {\bibfnamefont {M.}~\bibnamefont
  {Kawasaki}}, \ and\ \bibinfo {author} {\bibfnamefont {Y.}~\bibnamefont
  {Tokura}},\ }\bibfield  {title} {\enquote {\bibinfo {title} {A magnetic
  heterostructure of topological insulators as a candidate for an axion
  insulator}}, }\href {\doibase 10.1038/nmat4855} {\bibfield  {journal}
  {\bibinfo  {journal} {Nat. Mater.}\ }\textbf {\bibinfo {volume} {16}},\
  \bibinfo {pages} {516} (\bibinfo {year} {2017}{\natexlab{a}})}\BibitemShut
  {NoStop}%
\bibitem [{\citenamefont {Mogi}\ \emph
  {et~al.}(2017{\natexlab{b}})\citenamefont {Mogi}, \citenamefont {Kawamura},
  \citenamefont {Tsukazaki}, \citenamefont {Yoshimi}, \citenamefont
  {Takahashi}, \citenamefont {Kawasaki},\ and\ \citenamefont
  {Tokura}}]{Mogi17sa}%
  \BibitemOpen
  \bibfield  {author} {\bibinfo {author} {\bibfnamefont {M.}~\bibnamefont
  {Mogi}}, \bibinfo {author} {\bibfnamefont {M.}~\bibnamefont {Kawamura}},
  \bibinfo {author} {\bibfnamefont {A.}~\bibnamefont {Tsukazaki}}, \bibinfo
  {author} {\bibfnamefont {R.}~\bibnamefont {Yoshimi}}, \bibinfo {author}
  {\bibfnamefont {K.~S.}\ \bibnamefont {Takahashi}}, \bibinfo {author}
  {\bibfnamefont {M.}~\bibnamefont {Kawasaki}}, \ and\ \bibinfo {author}
  {\bibfnamefont {Y.}~\bibnamefont {Tokura}},\ }\bibfield  {title} {\enquote
  {\bibinfo {title} {Tailoring tricolor structure of magnetic topological
  insulator for robust axion insulator}}, }\href {\doibase
  10.1126/sciadv.aao1669} {\bibfield  {journal} {\bibinfo  {journal} {Sci.
  Adv.}\ }\textbf {\bibinfo {volume} {3}},\ \bibinfo {pages} {eaao1669}
  (\bibinfo {year} {2017}{\natexlab{b}})}\BibitemShut {NoStop}%
\bibitem [{\citenamefont {Deng}\ \emph {et~al.}(2020)\citenamefont {Deng},
  \citenamefont {Yu}, \citenamefont {Shi}, \citenamefont {Guo}, \citenamefont
  {Xu}, \citenamefont {Wang}, \citenamefont {Chen},\ and\ \citenamefont
  {Zhang}}]{Deng20sci}%
  \BibitemOpen
  \bibfield  {author} {\bibinfo {author} {\bibfnamefont {Y.}~\bibnamefont
  {Deng}}, \bibinfo {author} {\bibfnamefont {Y.}~\bibnamefont {Yu}}, \bibinfo
  {author} {\bibfnamefont {M.~Z.}\ \bibnamefont {Shi}}, \bibinfo {author}
  {\bibfnamefont {Z.}~\bibnamefont {Guo}}, \bibinfo {author} {\bibfnamefont
  {Z.}~\bibnamefont {Xu}}, \bibinfo {author} {\bibfnamefont {J.}~\bibnamefont
  {Wang}}, \bibinfo {author} {\bibfnamefont {X.~H.}\ \bibnamefont {Chen}}, \
  and\ \bibinfo {author} {\bibfnamefont {Y.}~\bibnamefont {Zhang}},\ }\bibfield
   {title} {\enquote {\bibinfo {title} {Quantum anomalous {Hall} effect in
  intrinsic magnetic topological insulator {MnBi}$_2${Te}$_4$}}, }\href
  {https://science.sciencemag.org/content/367/6480/895} {\bibfield  {journal}
  {\bibinfo  {journal} {Science}\ }\textbf {\bibinfo {volume} {367}},\ \bibinfo
  {pages} {895} (\bibinfo {year} {2020})}\BibitemShut {NoStop}%
\bibitem [{\citenamefont {Gao}\ \emph {et~al.}(2021)\citenamefont {Gao},
  \citenamefont {Liu}, \citenamefont {Hu}, \citenamefont {Qiu}, \citenamefont
  {Tzschaschel}, \citenamefont {Ghosh}, \citenamefont {Ho}, \citenamefont
  {B{\'{e}}rub{\'{e}}}, \citenamefont {Chen}, \citenamefont {Sun},
  \citenamefont {Zhang}, \citenamefont {Zhang}, \citenamefont {Wang},
  \citenamefont {Wang}, \citenamefont {Huang}, \citenamefont {Felser},
  \citenamefont {Agarwal}, \citenamefont {Ding}, \citenamefont {Tien},
  \citenamefont {Akey}, \citenamefont {Gardener}, \citenamefont {Singh},
  \citenamefont {Watanabe}, \citenamefont {Taniguchi}, \citenamefont {Burch},
  \citenamefont {Bell}, \citenamefont {Zhou}, \citenamefont {Gao},
  \citenamefont {Lu}, \citenamefont {Bansil}, \citenamefont {Lin},
  \citenamefont {Chang}, \citenamefont {Fu}, \citenamefont {Ma}, \citenamefont
  {Ni},\ and\ \citenamefont {Xu}}]{Gao2021Nature}%
  \BibitemOpen
  \bibfield  {author} {\bibinfo {author} {\bibfnamefont {A.}~\bibnamefont
  {Gao}}, \bibinfo {author} {\bibfnamefont {Y.-F.}\ \bibnamefont {Liu}},
  \bibinfo {author} {\bibfnamefont {C.}~\bibnamefont {Hu}}, \bibinfo {author}
  {\bibfnamefont {J.-X.}\ \bibnamefont {Qiu}}, \bibinfo {author} {\bibfnamefont
  {C.}~\bibnamefont {Tzschaschel}}, \bibinfo {author} {\bibfnamefont
  {B.}~\bibnamefont {Ghosh}},  \emph {et~al.},\ }\bibfield  {title} {\enquote
  {\bibinfo {title} {{Layer Hall effect in a 2D topological axion
  antiferromagnet}}}, }\href {\doibase 10.1038/s41586-021-03679-w} {\bibfield
  {journal} {\bibinfo  {journal} {Nature}\ }\textbf {\bibinfo {volume} {595}},\
  \bibinfo {pages} {521} (\bibinfo {year} {2021})}\BibitemShut {NoStop}%
\bibitem [{\citenamefont {Chen}\ \emph
  {et~al.}(2021{\natexlab{b}})\citenamefont {Chen}, \citenamefont {Li},
  \citenamefont {Sun}, \citenamefont {Liu}, \citenamefont {Zhao}, \citenamefont
  {Lu},\ and\ \citenamefont {Xie}}]{ChenR21PRB}%
  \BibitemOpen
  \bibfield  {author} {\bibinfo {author} {\bibfnamefont {R.}~\bibnamefont
  {Chen}}, \bibinfo {author} {\bibfnamefont {S.}~\bibnamefont {Li}}, \bibinfo
  {author} {\bibfnamefont {H.-P.}\ \bibnamefont {Sun}}, \bibinfo {author}
  {\bibfnamefont {Q.}~\bibnamefont {Liu}}, \bibinfo {author} {\bibfnamefont
  {Y.}~\bibnamefont {Zhao}}, \bibinfo {author} {\bibfnamefont {H.-Z.}\
  \bibnamefont {Lu}}, \ and\ \bibinfo {author} {\bibfnamefont {X.~C.}\
  \bibnamefont {Xie}},\ }\bibfield  {title} {\enquote {\bibinfo {title} {Using
  nonlocal surface transport to identify the axion insulator}}, }\href
  {\doibase 10.1103/PhysRevB.103.L241409} {\bibfield  {journal} {\bibinfo
  {journal} {Phys. Rev. B}\ }\textbf {\bibinfo {volume} {103}},\ \bibinfo
  {pages} {L241409} (\bibinfo {year} {2021}{\natexlab{b}})}\BibitemShut
  {NoStop}%
\bibitem [{\citenamefont {Ding}\ \emph {et~al.}(2020)\citenamefont {Ding},
  \citenamefont {Xu}, \citenamefont {Chen},\ and\ \citenamefont
  {Xie}}]{Ding20prbrc}%
  \BibitemOpen
  \bibfield  {author} {\bibinfo {author} {\bibfnamefont {Y.-R.}\ \bibnamefont
  {Ding}}, \bibinfo {author} {\bibfnamefont {D.-H.}\ \bibnamefont {Xu}},
  \bibinfo {author} {\bibfnamefont {C.-Z.}\ \bibnamefont {Chen}}, \ and\
  \bibinfo {author} {\bibfnamefont {X.~C.}\ \bibnamefont {Xie}},\ }\bibfield
  {title} {\enquote {\bibinfo {title} {{Hinged quantum spin Hall effect in
  antiferromagnetic topological insulators}}}, }\href {\doibase
  10.1103/physrevb.101.041404} {\bibfield  {journal} {\bibinfo  {journal}
  {Phys. Rev. B}\ }\textbf {\bibinfo {volume} {101}},\ \bibinfo {pages}
  {041404(R)} (\bibinfo {year} {2020})}\BibitemShut {NoStop}%
\bibitem [{\citenamefont {Shumiya}\ \emph {et~al.}(2022)\citenamefont
  {Shumiya}, \citenamefont {Hossain}, \citenamefont {Yin}, \citenamefont
  {Wang}, \citenamefont {Litskevich}, \citenamefont {Yoon}, \citenamefont {Li},
  \citenamefont {Yang}, \citenamefont {Jiang}, \citenamefont {Cheng},
  \citenamefont {Lin}, \citenamefont {Zhang}, \citenamefont {Cheng},
  \citenamefont {Cochran}, \citenamefont {Multer}, \citenamefont {Yang},
  \citenamefont {Casas}, \citenamefont {Chang}, \citenamefont {Neupert},
  \citenamefont {Yuan}, \citenamefont {Jia}, \citenamefont {Lin}, \citenamefont
  {Yao}, \citenamefont {Balicas}, \citenamefont {Zhang}, \citenamefont {Yao},\
  and\ \citenamefont {Hasan}}]{Shumiya2022NatMat}%
  \BibitemOpen
  \bibfield  {author} {\bibinfo {author} {\bibfnamefont {N.}~\bibnamefont
  {Shumiya}}, \bibinfo {author} {\bibfnamefont {M.~S.}\ \bibnamefont
  {Hossain}}, \bibinfo {author} {\bibfnamefont {J.-X.}\ \bibnamefont {Yin}},
  \bibinfo {author} {\bibfnamefont {Z.}~\bibnamefont {Wang}}, \bibinfo {author}
  {\bibfnamefont {M.}~\bibnamefont {Litskevich}}, \bibinfo {author}
  {\bibfnamefont {C.}~\bibnamefont {Yoon}},  \emph {et~al.},\ }\bibfield
  {title} {\enquote {\bibinfo {title} {Evidence of a room-temperature quantum
  spin hall edge state in a higher-order topological insulator}}, }\href
  {\doibase 10.1038/s41563-022-01304-3} {\bibfield  {journal} {\bibinfo
  {journal} {Nat. Mat.}\ }\textbf {\bibinfo {volume} {21}},\ \bibinfo {pages}
  {1111} (\bibinfo {year} {2022})}\BibitemShut {NoStop}%
\bibitem [{\citenamefont {Zhao}\ \emph {et~al.}(2020)\citenamefont {Zhao},
  \citenamefont {Zhang}, \citenamefont {Mei}, \citenamefont {Zhou},
  \citenamefont {Yi}, \citenamefont {Zhang}, \citenamefont {Yu}, \citenamefont
  {Xiao}, \citenamefont {Wang}, \citenamefont {Samarth}, \citenamefont {Chan},
  \citenamefont {Liu},\ and\ \citenamefont {Chang}}]{Zhao2020Nature}%
  \BibitemOpen
  \bibfield  {author} {\bibinfo {author} {\bibfnamefont {Y.-F.}\ \bibnamefont
  {Zhao}}, \bibinfo {author} {\bibfnamefont {R.}~\bibnamefont {Zhang}},
  \bibinfo {author} {\bibfnamefont {R.}~\bibnamefont {Mei}}, \bibinfo {author}
  {\bibfnamefont {L.-J.}\ \bibnamefont {Zhou}}, \bibinfo {author}
  {\bibfnamefont {H.}~\bibnamefont {Yi}}, \bibinfo {author} {\bibfnamefont
  {Y.-Q.}\ \bibnamefont {Zhang}},  \emph {et~al.},\ }\bibfield  {title}
  {\enquote {\bibinfo {title} {{Tuning the Chern number in quantum anomalous
  Hall insulators}}}, }\href {\doibase 10.1038/s41586-020-3020-3} {\bibfield
  {journal} {\bibinfo  {journal} {Nature}\ }\textbf {\bibinfo {volume} {588}},\
  \bibinfo {pages} {419} (\bibinfo {year} {2020})}\BibitemShut {NoStop}%
\bibitem [{\citenamefont {Agarwala}\ \emph {et~al.}(2020)\citenamefont
  {Agarwala}, \citenamefont {Juri\ifmmode \check{c}\else
  \v{c}\fi{}i\ifmmode~\acute{c}\else \'{c}\fi{}},\ and\ \citenamefont
  {Roy}}]{Agarwala2020PRR}%
  \BibitemOpen
  \bibfield  {author} {\bibinfo {author} {\bibfnamefont {A.}~\bibnamefont
  {Agarwala}}, \bibinfo {author} {\bibfnamefont {V.}~\bibnamefont {Juri\ifmmode
  \check{c}\else \v{c}\fi{}i\ifmmode~\acute{c}\else \'{c}\fi{}}}, \ and\
  \bibinfo {author} {\bibfnamefont {B.}~\bibnamefont {Roy}},\ }\bibfield
  {title} {\enquote {\bibinfo {title} {Higher-order topological insulators in
  amorphous solids}}, }\href {\doibase 10.1103/PhysRevResearch.2.012067}
  {\bibfield  {journal} {\bibinfo  {journal} {Phys. Rev. Res.}\ }\textbf
  {\bibinfo {volume} {2}},\ \bibinfo {pages} {012067} (\bibinfo {year}
  {2020})}\BibitemShut {NoStop}%
\bibitem [{\citenamefont {Wang}\ and\ \citenamefont
  {Moore}(2019)}]{WangY2019PRB}%
  \BibitemOpen
  \bibfield  {author} {\bibinfo {author} {\bibfnamefont {Y.-Q.}\ \bibnamefont
  {Wang}}\ and\ \bibinfo {author} {\bibfnamefont {J.~E.}\ \bibnamefont
  {Moore}},\ }\bibfield  {title} {\enquote {\bibinfo {title} {Boundary edge
  networks induced by bulk topology}}, }\href {\doibase
  10.1103/PhysRevB.99.155102} {\bibfield  {journal} {\bibinfo  {journal} {Phys.
  Rev. B}\ }\textbf {\bibinfo {volume} {99}},\ \bibinfo {pages} {155102}
  (\bibinfo {year} {2019})}\BibitemShut {NoStop}%
\bibitem [{\citenamefont {Tao}\ and\ \citenamefont {Xu}(2022)}]{Tao2022arXiv}%
  \BibitemOpen
  \bibfield  {author} {\bibinfo {author} {\bibfnamefont {Y.-L.}\ \bibnamefont
  {Tao}}\ and\ \bibinfo {author} {\bibfnamefont {Y.}~\bibnamefont {Xu}},\
  }\bibfield  {title} {\enquote {\bibinfo {title} {Higher-order topological
  hyperbolic lattices}}, }\href {https://arxiv.org/pdf/2209.02262.pdf}
  {\bibfield  {journal} {\bibinfo  {journal} {arXiv:2209.02262}\ } (\bibinfo
  {year} {2022})}\BibitemShut {NoStop}%
\bibitem [{\citenamefont {Roy}(2019)}]{PRR2019Roy}%
  \BibitemOpen
  \bibfield  {author} {\bibinfo {author} {\bibfnamefont {B.}~\bibnamefont
  {Roy}},\ }\bibfield  {title} {\enquote {\bibinfo {title} {Antiunitary
  symmetry protected higher-order topological phases}}, }\href {\doibase
  10.1103/PhysRevResearch.1.032048} {\bibfield  {journal} {\bibinfo  {journal}
  {Phys. Rev. Res.}\ }\textbf {\bibinfo {volume} {1}},\ \bibinfo {pages}
  {032048} (\bibinfo {year} {2019})}\BibitemShut {NoStop}%
\bibitem [{\citenamefont {Pixley}\ \emph {et~al.}(2016)\citenamefont {Pixley},
  \citenamefont {Huse},\ and\ \citenamefont {Das~Sarma}}]{Pixley2016PRX}%
  \BibitemOpen
  \bibfield  {author} {\bibinfo {author} {\bibfnamefont {J.~H.}\ \bibnamefont
  {Pixley}}, \bibinfo {author} {\bibfnamefont {D.~A.}\ \bibnamefont {Huse}}, \
  and\ \bibinfo {author} {\bibfnamefont {S.}~\bibnamefont {Das~Sarma}},\
  }\bibfield  {title} {\enquote {\bibinfo {title} {Rare-region-induced avoided
  quantum criticality in disordered three-dimensional {Dirac} and {Weyl}
  semimetals}}, }\href {\doibase 10.1103/PhysRevX.6.021042} {\bibfield
  {journal} {\bibinfo  {journal} {Phys. Rev. X}\ }\textbf {\bibinfo {volume}
  {6}},\ \bibinfo {pages} {021042} (\bibinfo {year} {2016})}\BibitemShut
  {NoStop}%
\bibitem [{\citenamefont {Roy}\ \emph {et~al.}(2018)\citenamefont {Roy},
  \citenamefont {Slager},\ and\ \citenamefont
  {Juri{\v{c}}i{\'{c}}}}]{Roy2018PRX}%
  \BibitemOpen
  \bibfield  {author} {\bibinfo {author} {\bibfnamefont {B.}~\bibnamefont
  {Roy}}, \bibinfo {author} {\bibfnamefont {R.-J.}\ \bibnamefont {Slager}}, \
  and\ \bibinfo {author} {\bibfnamefont {V.}~\bibnamefont
  {Juri{\v{c}}i{\'{c}}}},\ }\bibfield  {title} {\enquote {\bibinfo {title}
  {Global phase diagram of a dirty weyl liquid and emergent
  superuniversality}}, }\href {\doibase 10.1103/physrevx.8.031076} {\bibfield
  {journal} {\bibinfo  {journal} {Phys. Rev. X}\ }\textbf {\bibinfo {volume}
  {8}},\ \bibinfo {pages} {031076} (\bibinfo {year} {2018})}\BibitemShut
  {NoStop}%
\bibitem [{\citenamefont {Benalcazar}\ \emph {et~al.}(2019)\citenamefont
  {Benalcazar}, \citenamefont {Li},\ and\ \citenamefont
  {Hughes}}]{Benalcazar19PRB}%
  \BibitemOpen
  \bibfield  {author} {\bibinfo {author} {\bibfnamefont {W.~A.}\ \bibnamefont
  {Benalcazar}}, \bibinfo {author} {\bibfnamefont {T.}~\bibnamefont {Li}}, \
  and\ \bibinfo {author} {\bibfnamefont {T.~L.}\ \bibnamefont {Hughes}},\
  }\bibfield  {title} {\enquote {\bibinfo {title} {Quantization of fractional
  corner charge in ${C}_{n}$-symmetric higher-order topological crystalline
  insulators}}, }\href {\doibase 10.1103/PhysRevB.99.245151} {\bibfield
  {journal} {\bibinfo  {journal} {Phys. Rev. B}\ }\textbf {\bibinfo {volume}
  {99}},\ \bibinfo {pages} {245151} (\bibinfo {year} {2019})}\BibitemShut
  {NoStop}%
\bibitem [{\citenamefont {van Miert}\ and\ \citenamefont
  {Ortix}(2018)}]{Miert18PRB}%
  \BibitemOpen
  \bibfield  {author} {\bibinfo {author} {\bibfnamefont {G.}~\bibnamefont {van
  Miert}}\ and\ \bibinfo {author} {\bibfnamefont {C.}~\bibnamefont {Ortix}},\
  }\bibfield  {title} {\enquote {\bibinfo {title} {Dislocation charges reveal
  two-dimensional topological crystalline invariants}}, }\href {\doibase
  10.1103/PhysRevB.97.201111} {\bibfield  {journal} {\bibinfo  {journal} {Phys.
  Rev. B}\ }\textbf {\bibinfo {volume} {97}},\ \bibinfo {pages} {201111}
  (\bibinfo {year} {2018})}\BibitemShut {NoStop}%
\bibitem [{\citenamefont {Geier}\ \emph {et~al.}(2021)\citenamefont {Geier},
  \citenamefont {Fulga},\ and\ \citenamefont {Lau}}]{Geier2021SPP}%
  \BibitemOpen
  \bibfield  {author} {\bibinfo {author} {\bibfnamefont {M.}~\bibnamefont
  {Geier}}, \bibinfo {author} {\bibfnamefont {I.~C.}\ \bibnamefont {Fulga}}, \
  and\ \bibinfo {author} {\bibfnamefont {A.}~\bibnamefont {Lau}},\ }\bibfield
  {title} {\enquote {\bibinfo {title} {Bulk-boundary-defect correspondence at
  disclinations in rotation-symmetric topological insulators and
  superconductors}}, }\href {\doibase 10.21468/scipostphys.10.4.092} {\bibfield
   {journal} {\bibinfo  {journal} {{SciPost} Phys.}\ }\textbf {\bibinfo
  {volume} {10}},\ \bibinfo {pages} {092} (\bibinfo {year} {2021})}\BibitemShut
  {NoStop}%
\bibitem [{\citenamefont {Qi}\ \emph {et~al.}(2022)\citenamefont {Qi},
  \citenamefont {He},\ and\ \citenamefont {Xiao}}]{Qi2022APL}%
  \BibitemOpen
  \bibfield  {author} {\bibinfo {author} {\bibfnamefont {Y.}~\bibnamefont
  {Qi}}, \bibinfo {author} {\bibfnamefont {H.}~\bibnamefont {He}}, \ and\
  \bibinfo {author} {\bibfnamefont {M.}~\bibnamefont {Xiao}},\ }\bibfield
  {title} {\enquote {\bibinfo {title} {Manipulation of acoustic vortex with
  topological dislocation states}}, }\href {\doibase 10.1063/5.0095543}
  {\bibfield  {journal} {\bibinfo  {journal} {Appl. Phys. Lett.}\ }\textbf
  {\bibinfo {volume} {120}},\ \bibinfo {pages} {212202} (\bibinfo {year}
  {2022})}\BibitemShut {NoStop}%
\bibitem [{\citenamefont {Peterson}\ \emph {et~al.}(2021)\citenamefont
  {Peterson}, \citenamefont {Li}, \citenamefont {Jiang}, \citenamefont
  {Hughes},\ and\ \citenamefont {Bahl}}]{Peterson2021Nature}%
  \BibitemOpen
  \bibfield  {author} {\bibinfo {author} {\bibfnamefont {C.~W.}\ \bibnamefont
  {Peterson}}, \bibinfo {author} {\bibfnamefont {T.}~\bibnamefont {Li}},
  \bibinfo {author} {\bibfnamefont {W.}~\bibnamefont {Jiang}}, \bibinfo
  {author} {\bibfnamefont {T.~L.}\ \bibnamefont {Hughes}}, \ and\ \bibinfo
  {author} {\bibfnamefont {G.}~\bibnamefont {Bahl}},\ }\bibfield  {title}
  {\enquote {\bibinfo {title} {Trapped fractional charges at bulk defects in
  topological insulators}}, }\href {\doibase 10.1038/s41586-020-03117-3}
  {\bibfield  {journal} {\bibinfo  {journal} {Nature}\ }\textbf {\bibinfo
  {volume} {589}},\ \bibinfo {pages} {376} (\bibinfo {year}
  {2021})}\BibitemShut {NoStop}%
\bibitem [{\citenamefont {Ye}\ \emph {et~al.}(2022)\citenamefont {Ye},
  \citenamefont {Qiu}, \citenamefont {Xiao}, \citenamefont {Li}, \citenamefont
  {Du}, \citenamefont {Ke},\ and\ \citenamefont {Liu}}]{Ye2021NC}%
  \BibitemOpen
  \bibfield  {author} {\bibinfo {author} {\bibfnamefont {L.}~\bibnamefont
  {Ye}}, \bibinfo {author} {\bibfnamefont {C.}~\bibnamefont {Qiu}}, \bibinfo
  {author} {\bibfnamefont {M.}~\bibnamefont {Xiao}}, \bibinfo {author}
  {\bibfnamefont {T.}~\bibnamefont {Li}}, \bibinfo {author} {\bibfnamefont
  {J.}~\bibnamefont {Du}}, \bibinfo {author} {\bibfnamefont {M.}~\bibnamefont
  {Ke}}, \ and\ \bibinfo {author} {\bibfnamefont {Z.}~\bibnamefont {Liu}},\
  }\bibfield  {title} {\enquote {\bibinfo {title} {Topological dislocation
  modes in three-dimensional acoustic topological insulators}}, }\href
  {\doibase 10.1038/s41467-022-28182-2} {\bibfield  {journal} {\bibinfo
  {journal} {Nat. Commu.}\ }\textbf {\bibinfo {volume} {13}},\ \bibinfo {pages}
  {508} (\bibinfo {year} {2022})}\BibitemShut {NoStop}%
\bibitem [{\citenamefont {Xue}\ \emph {et~al.}(2021)\citenamefont {Xue},
  \citenamefont {Jia}, \citenamefont {Ge}, \citenamefont {Guan}, \citenamefont
  {Wang}, \citenamefont {Yuan}, \citenamefont {Sun}, \citenamefont {Chong},\
  and\ \citenamefont {Zhang}}]{Xue21PRL}%
  \BibitemOpen
  \bibfield  {author} {\bibinfo {author} {\bibfnamefont {H.}~\bibnamefont
  {Xue}}, \bibinfo {author} {\bibfnamefont {D.}~\bibnamefont {Jia}}, \bibinfo
  {author} {\bibfnamefont {Y.}~\bibnamefont {Ge}}, \bibinfo {author}
  {\bibfnamefont {Y.-j.}\ \bibnamefont {Guan}}, \bibinfo {author}
  {\bibfnamefont {Q.}~\bibnamefont {Wang}}, \bibinfo {author} {\bibfnamefont
  {S.-q.}\ \bibnamefont {Yuan}}, \bibinfo {author} {\bibfnamefont {H.-x.}\
  \bibnamefont {Sun}}, \bibinfo {author} {\bibfnamefont {Y.~D.}\ \bibnamefont
  {Chong}}, \ and\ \bibinfo {author} {\bibfnamefont {B.}~\bibnamefont
  {Zhang}},\ }\bibfield  {title} {\enquote {\bibinfo {title} {Observation of
  dislocation-induced topological modes in a three-dimensional acoustic
  topological insulator}}, }\href {\doibase 10.1103/PhysRevLett.127.214301}
  {\bibfield  {journal} {\bibinfo  {journal} {Phys. Rev. Lett.}\ }\textbf
  {\bibinfo {volume} {127}},\ \bibinfo {pages} {214301} (\bibinfo {year}
  {2021})}\BibitemShut {NoStop}%
\bibitem [{\citenamefont {Deng}\ \emph {et~al.}(2022)\citenamefont {Deng},
  \citenamefont {Benalcazar}, \citenamefont {Chen}, \citenamefont {Oudich},
  \citenamefont {Ma},\ and\ \citenamefont {Jing}}]{Deng22PRL}%
  \BibitemOpen
  \bibfield  {author} {\bibinfo {author} {\bibfnamefont {Y.}~\bibnamefont
  {Deng}}, \bibinfo {author} {\bibfnamefont {W.~A.}\ \bibnamefont
  {Benalcazar}}, \bibinfo {author} {\bibfnamefont {Z.-G.}\ \bibnamefont
  {Chen}}, \bibinfo {author} {\bibfnamefont {M.}~\bibnamefont {Oudich}},
  \bibinfo {author} {\bibfnamefont {G.}~\bibnamefont {Ma}}, \ and\ \bibinfo
  {author} {\bibfnamefont {Y.}~\bibnamefont {Jing}},\ }\bibfield  {title}
  {\enquote {\bibinfo {title} {Observation of degenerate zero-energy
  topological states at disclinations in an acoustic lattice}}, }\href
  {\doibase 10.1103/PhysRevLett.128.174301} {\bibfield  {journal} {\bibinfo
  {journal} {Phys. Rev. Lett.}\ }\textbf {\bibinfo {volume} {128}},\ \bibinfo
  {pages} {174301} (\bibinfo {year} {2022})}\BibitemShut {NoStop}%
\bibitem [{\citenamefont {Xia}\ \emph {et~al.}(2022)\citenamefont {Xia},
  \citenamefont {Jiang}, \citenamefont {Tong}, \citenamefont {Zheng},\ and\
  \citenamefont {Man}}]{Xia2022AMS}%
  \BibitemOpen
  \bibfield  {author} {\bibinfo {author} {\bibfnamefont {B.}~\bibnamefont
  {Xia}}, \bibinfo {author} {\bibfnamefont {Z.}~\bibnamefont {Jiang}}, \bibinfo
  {author} {\bibfnamefont {L.}~\bibnamefont {Tong}}, \bibinfo {author}
  {\bibfnamefont {S.}~\bibnamefont {Zheng}}, \ and\ \bibinfo {author}
  {\bibfnamefont {X.}~\bibnamefont {Man}},\ }\bibfield  {title} {\enquote
  {\bibinfo {title} {Topological bound states in elastic phononic plates
  induced by disclinations}}, }\href {\doibase 10.1007/s10409-021-09083-0}
  {\bibfield  {journal} {\bibinfo  {journal} {Acta Mech. Sin.}\ }\textbf
  {\bibinfo {volume} {38}},\ \bibinfo {pages} {521459} (\bibinfo {year}
  {2022})}\BibitemShut {NoStop}%
\bibitem [{\citenamefont {Wang}\ \emph {et~al.}(2021)\citenamefont {Wang},
  \citenamefont {Ge}, \citenamefont {xiang Sun}, \citenamefont {Xue},
  \citenamefont {Jia}, \citenamefont {jun Guan}, \citenamefont {qi~Yuan},
  \citenamefont {Zhang},\ and\ \citenamefont {Chong}}]{WangQ2021NC}%
  \BibitemOpen
  \bibfield  {author} {\bibinfo {author} {\bibfnamefont {Q.}~\bibnamefont
  {Wang}}, \bibinfo {author} {\bibfnamefont {Y.}~\bibnamefont {Ge}}, \bibinfo
  {author} {\bibfnamefont {H.}~\bibnamefont {xiang Sun}}, \bibinfo {author}
  {\bibfnamefont {H.}~\bibnamefont {Xue}}, \bibinfo {author} {\bibfnamefont
  {D.}~\bibnamefont {Jia}}, \bibinfo {author} {\bibfnamefont {Y.}~\bibnamefont
  {jun Guan}}, \bibinfo {author} {\bibfnamefont {S.}~\bibnamefont {qi~Yuan}},
  \bibinfo {author} {\bibfnamefont {B.}~\bibnamefont {Zhang}}, \ and\ \bibinfo
  {author} {\bibfnamefont {Y.~D.}\ \bibnamefont {Chong}},\ }\bibfield  {title}
  {\enquote {\bibinfo {title} {Vortex states in an acoustic weyl crystal with a
  topological lattice defect}}, }\href {\doibase 10.1038/s41467-021-23963-7}
  {\bibfield  {journal} {\bibinfo  {journal} {Nat. Commun.}\ }\textbf {\bibinfo
  {volume} {12}},\ \bibinfo {pages} {3654} (\bibinfo {year}
  {2021})}\BibitemShut {NoStop}%
\bibitem [{\citenamefont {Kleman}\ and\ \citenamefont
  {Friedel}(2008)}]{Kleman08RMP}%
  \BibitemOpen
  \bibfield  {author} {\bibinfo {author} {\bibfnamefont {M.}~\bibnamefont
  {Kleman}}\ and\ \bibinfo {author} {\bibfnamefont {J.}~\bibnamefont
  {Friedel}},\ }\bibfield  {title} {\enquote {\bibinfo {title} {Disclinations,
  dislocations, and continuous defects: A reappraisal}}, }\href {\doibase
  10.1103/RevModPhys.80.61} {\bibfield  {journal} {\bibinfo  {journal} {Rev.
  Mod. Phys.}\ }\textbf {\bibinfo {volume} {80}},\ \bibinfo {pages} {61}
  (\bibinfo {year} {2008})}\BibitemShut {NoStop}%
\end{thebibliography}%


\end{document}